\documentclass[12pt]{article}
\pdfoutput=1
\usepackage{graphicx,psfrag,epsf,color}
\usepackage{xcolor}
\usepackage{amsmath,amssymb,amsfonts}
\usepackage{appendix}
\usepackage{multirow}
\usepackage{geometry}
\usepackage{float}
\usepackage{mathrsfs}  
\usepackage{authblk}
\usepackage{array}
\usepackage{cite}
\usepackage{slashed}
\usepackage{graphicx}
\usepackage{rotating}
\usepackage{slashed, cancel}
\usepackage[caption=false]{subfig}
\newcounter{MBQ}

\newcounter{MBQQ}

\usepackage{tabularx}
\newcolumntype{C}{>{\centering\arraybackslash}X}

\usepackage{bigints}

\newcommand{\eps}{\epsilon}

\newcommand{\mut}{\tilde{\mu}}
\def\be{\begin{equation}}
\def\ee{\end{equation}}
\def\beq{\begin{eqnarray}}
\def\eeq{\end{eqnarray}}
\usepackage{hyperref}
\newcommand{\bea}{\begin{eqnarray}}
\newcommand{\eea}{\end{eqnarray}}
\newcommand{\beas}{\begin{eqnarray*}}
\newcommand{\eeas}{\end{eqnarray*}}

\newcommand{\aqed}{a}
\newcommand{\Vqed}{V}

\newcommand{\qboth}{\otimes}

\newcommand{\alem}{\alpha_{\rm em}}

\newcommand{\nn}{\nonumber}

\newcommand{\bra}[1]{\big\langle{#1}\big\vert}
\newcommand{\ket}[1]{\big\vert{#1}\big\rangle}

\numberwithin{equation}{section}

\begin{document}
\allowdisplaybreaks

\begin{titlepage}

\begin{flushright}
{\small
TUM-HEP-1397/22\\
MITP-21-062\\
Nikhef-2022-004\\
April 19, 2022 \\
}
\end{flushright}

\vskip0.8cm
\begin{center}
{\Large \bf\boldmath Light-cone distribution amplitudes of heavy 
mesons\\[0.15cm] with QED effects}
\end{center}

\vspace{0.5cm}
\begin{center}
{\sc Martin~Beneke,$^a$ Philipp~B\"oer,$^b$ 
Jan-Niklas Toelstede,$^{a}$ K. Keri Vos$^{c,d}$} \\[6mm]
{\it $^a$Physik Department T31,\\
James-Franck-Stra\ss{}e~1, 
Technische Universit\"at M\"unchen,\\
D--85748 Garching, Germany\\[0.3cm]

{\it $^b$PRISMA$^+$ Cluster of Excellence \& Mainz Institute for Theoretical Physics,\\ 
Staudingerweg 9, Johannes Gutenberg University,\\ 
D-55128 Mainz, Germany}\\[0.3cm]

{\it $^c$Gravitational 
Waves and Fundamental Physics (GWFP),\\ 
Maastricht University, Duboisdomein 30,\\ 
NL-6229 GT Maastricht, the
Netherlands}\\[0.3cm]

{\it $^d$Nikhef, Science Park 105,\\ 
NL-1098 XG Amsterdam, the Netherlands}}
\end{center}

\vspace{0.6cm}
\begin{abstract}
\vskip0.2cm\noindent
We discuss the QED-generalized leading-twist light-cone distribution amplitudes of heavy mesons, that appear in QCD$\times$QED factorization theorems for exclusive two-body $B$ decays. In the presence of electrically charged particles, these functions should be more appropriately regarded as soft functions for heavy-meson decays into two back-to-back particles. In this paper, we derive the one-loop anomalous dimension of these soft functions and study their behaviour under renormalization-scale evolution, obtaining an exact solution in Laplace space. In addition, we provide numerical solutions for the soft functions and analytical solutions to all orders in the strong and to first order in the electromagnetic coupling. For the inverse (and inverse-logarithmic) moments, we obtain an all-order solution in both couplings. We further provide numerical estimates for QED corrections to the inverse moments.
\end{abstract}
\end{titlepage}



\section{Introduction}
\label{sec:intro}
\label{sec:blcda} 
\label{sec:softfunctiondef}

In QCD, the leading-twist light-cone distribution amplitude (LCDA) $\phi^+_{B}(\omega;\mu)$ of a meson composed of a heavy quark and light anti-quark is defined by \cite{Grozin:1996pq,Beneke:2000wa} 
\begin{eqnarray}
\label{eq:BLCDAdefQCD}
\bra{0} \bar{q}_s(tn_-) [tn_-,0] \slashed{n}_-\gamma_5 h_v(0)\ket{\bar{B}_v}  &=& \, i F_{\rm stat}(\mu) \int_0^\infty d\omega \,e^{-i \omega t} \,
\phi_{B}^+(\omega;\mu) \,. 
\qquad
\end{eqnarray}
Here $n^\mu_-$ is a light-like reference vector and $[tn_-,0]$ a finite-distance Wilson line consisting of soft gluon fields. The $B$-meson LCDA was first used in the QCD factorization of spectator scattering in charmless, non-leptonic $B$ decays 
\cite{Beneke:1999br} and has turned 
out to be a crucial hadronic input to almost any exclusive $B$ 
decay to light, energetic particles. Although the LCDA measures the correlation between 
the constituents of the $B$ meson at light-like separation, which is probed in such 
decays, the matrix element (\ref{eq:BLCDAdefQCD}) captures
the non-perturbative soft fluctuations characteristic of the 
bound state. It is therefore defined in heavy-quark 
effective theory (HQET) in terms of the static heavy quark field 
$h_v$ and the heavy-quark mass independent heavy meson state 
$\ket{\bar{B}_v}$ of HQET. The scale-dependent static HQET meson decay constant is 
related to the scale-independent QCD $B$-meson decay 
constant by 
\begin{equation}
f_B = 
\frac{K(\mu) F_{\rm stat}(\mu)}{\sqrt{m_B}} \; ,
\end{equation}
where 
\begin{equation} \label{eq:decayRGE}
K(\mu) = 1+\frac{\alpha_s C_F}{4\pi} 
\left(3\ln\frac{m_b}{\mu}-2\right)+\mathcal{O}(\alpha_s^2)
\end{equation}
is a short-distance matching coefficient of the heavy-light 
current\cite{Ji:1991pr}. With this 
convention the zeroth moment integral $\int_0^\infty\,d\omega\,
\phi^+_{B}(\omega;\mu)$ is dimensionless. However, non-negative 
moments of $\phi_B(\omega;\mu)$ are divergent for large $\omega$.
Instead, the most important quantity in leading power factorization theorems is the first
inverse moment \cite{Beneke:1999br}
\begin{equation}
\label{eq:lambdaBdef}
\frac{1}{\lambda^+_B(\mu)} \equiv \int_0^\infty
\frac{d\omega}{\omega}\,\phi^+_{B}(\omega; \mu)
\end{equation}
and its logarithmic modifications, which have convergent 
integrals.

In this paper we consider the renormalization of the 
generalization to the $B$-meson LCDA when electromagnetic 
interactions are included. This generalization was introduced 
in~\cite{Beneke:2019slt} to calculate QED corrections to 
the $B$-decay $\bar{B}_q\to \ell^+\ell^-$ and further 
generalized in \cite{Beneke:2020vnb} to all possible 
electric charge combinations in charmless 
two-body $\bar{B}\to M_1 M_2$ decays. A qualitative and quantitative understanding of the QED-generalized LCDAs for heavy mesons is the last missing piece in the factorization of QED effects at scales higher than a few times $\Lambda_{\rm QCD}$ up to $m_B$. The QCD$\times$QED definition of these functions in two-particle decays is not as universal and process-independent as in QCD. Due to the non-decoupling 
of soft photons, the QED generalization of the $B$-meson LCDA retains
knowledge of the charges and directions of flight of 
the final-state particles through light-like Wilson lines. This is reflected in particular by the explicit appearance of soft rescattering phases, which entail major phenomenological modifications and motivate to consider the QED generalization as soft functions for the process rather than LCDAs in the conventional sense. For this reason, we will use the term \textit{soft function} instead of LCDA throughout the rest of this paper. The following definition applies to $B$ decays into final states that consist of two back-to-back charged particles with four-momentum aligned with the light-like vectors $n_-^\mu$ and $n_+^\mu$, satisfying $n_+\cdot n_-=2$. We define \cite{Beneke:2019slt,Beneke:2020vnb}
\begin{align}
\label{eq:BLCDAdef}
\frac{1}{R_c^{(Q_{M_1})}R_{\bar{c}}^{(Q_{M_2})}} &\bra{0} 
\bar{q}^{(q)}_s(tn_-) [tn_-,0]^{(q)}  
\,\slashed{n}_-\gamma_5 \, h_v(0) S_{n_-}^{\dagger (Q_{M_1})}(0) S_{n_+}^{\dagger (Q_{M_2})}(0) \ket{\bar{B}_v} \nonumber  \\ 
 &= \, i F_{\rm stat}(\mu) \int_{-\infty}^\infty d\omega\; e^{-i \omega t} \,\Phi_{B, \qboth}(\omega;\mu) \, ,
\end{align}
where the symbol $\qboth$ on $\Phi_{B, \qboth}(\omega;\mu)$ labels the four possible pairs $(Q_{M_1}, Q_{M_2})$ of electric charges $Q=\pm 1$ of the final-state mesons (or leptons) $M_1$ and $M_2$. Here $[tn_-,0]^{(q)}$ is the finite-distance Wilson line in QCD$\times$QED, and $S_{n\pm}$ are soft QED Wilson lines for outgoing charged particles.  Their definitions are given in Appendix~\ref{app:soft}. Note that the operator in~\eqref{eq:BLCDAdef} is invariant under $SU(3)\times U(1)_{\rm em}$ gauge transformations. In addition, $R_{c(\bar{c})}$ are rearrangement factors introduced in \cite{Beneke:2019slt,Beneke:2020vnb} in order to define a consistent renormalization group equation (RGE). When describing the full physical process these factors cancel with the corresponding rearrangement factors present in the matrix elements for the final-state particles, such as the light-meson LCDA discussed in \cite{Beneke:2021pkl}.  Finally, we note that in \eqref{eq:BLCDAdef} we choose to normalize the QCD$\times$QED soft functions to the static decay constant defined {\em in the absence} of QED. This definition has the advantage that all QED modifications are now contained in $\Phi_{B, \qboth}(\omega;\mu)$, and it is not necessary to define a QED generalization of the $B$-meson decay constant.   

The outline of this paper is as follows. In Sec.~\ref{sec:RGdef} we calculate the one-loop anomalous dimension of the soft functions $\Phi_{B,\qboth}$ and discuss crucial details of the computation. Furthermore, we derive the RGE for the first inverse and logarithmic moments from the anomalous dimension. For 
the evolution of these moments, we obtain an exact solution in Sec.~\ref{sec:analytic}. The RGE for the soft functions themselves is solved formally in Laplace space. We derive analytical expressions for the soft functions to all orders in the strong and to first order in the electromagnetic coupling for practical applications. We present numerical results in Sec.~\ref{sec:numerics} and conclude in Sec.~\ref{sec:conclusion}. In Appendices \ref{app:soft}--\ref{app:MeijerG} contain supplemental material that was used for calculations in the main text.


\section{Renormalization group equations}
\label{sec:RGdef}
In this section, we derive the RGE for the soft functions $\Phi_{B,\qboth}(\omega;\mu)$ and their first inverse and logarithmic moments by calculating the ultraviolet divergence of the one-loop corrections to the matrix element~\eqref{eq:BLCDAdef} in QCD$\times$QED shown in Fig.~\ref{fig:BLCDAdia}. To this end, we define the renormalization factor, including the external quark-field renormalization in the $\overline{\rm{MS}}$ scheme, by
\begin{equation}
\label{eq:B-Zfactor}
\mathcal{O}_{\qboth}^{\rm{ren}}(\omega;\mu) = \int_{-\infty}^\infty d\omega'~Z_\qboth(\omega,\omega';\mu)~\mathcal{O}^{\rm bare}_{\qboth}(\omega')\,.
\end{equation}
Here  $\mathcal{O}_\otimes(\omega)$ is the Fourier transform with 
respect to $t$ of the 
operator on the left-hand side of  \eqref{eq:BLCDAdef}, 
and we note that the soft rearrangement factors $R_{c(\bar{c})}$  in \eqref{eq:BLCDAdef} are part of the definition of the operator. The convolution in $\omega'$ reflects the fact that operators with different momentum variable $\omega$ can mix into each other. We emphasize that the integration over $\omega'$ in \eqref{eq:B-Zfactor} runs over the entire real axis, because, as will be seen below, some of the soft functions have support for $-\infty < \omega < \infty$. This is opposed to the QCD-only case, where the support of the LCDA is restricted to $\omega>0$ \cite{Lange:2003ff}. To obtain the anomalous dimension for $\Phi_{B,\qboth}(\omega;\mu)$, we need to compensate for the $\mu$-dependent decay constant $F_{\rm stat}(\mu)$ using \eqref{eq:decayRGE}. The anomalous dimension is given by
\begin{equation}
\label{eq:Gdef}
    \Gamma_\qboth(\omega, \omega'; \mu) = - \int_{-\infty}^\infty d\hat{\omega}~\frac{dZ_\qboth^{}(\omega,\hat{\omega};\mu)}{d\ln\mu} Z^{-1}_\qboth(\hat{\omega},\omega';\mu) +\delta(\omega-\omega') \frac{d F_{\rm stat}(\mu)}{d \ln \mu}    
\end{equation}
and the soft functions obey the RGE
\begin{equation}
\label{eq:B-RGE}
    \frac{d}{d\ln\mu}\Phi_{B,\qboth} (\omega; \mu) = - \int_{-\infty}^\infty d\omega'~\Gamma_\qboth(\omega,\omega';\mu)\Phi_{B,\qboth}(\omega'; \mu) \, .
\end{equation} 

\subsection{Details of the calculation}\label{sec:detcal}
\begin{figure}
\centering
\includegraphics[scale=1]{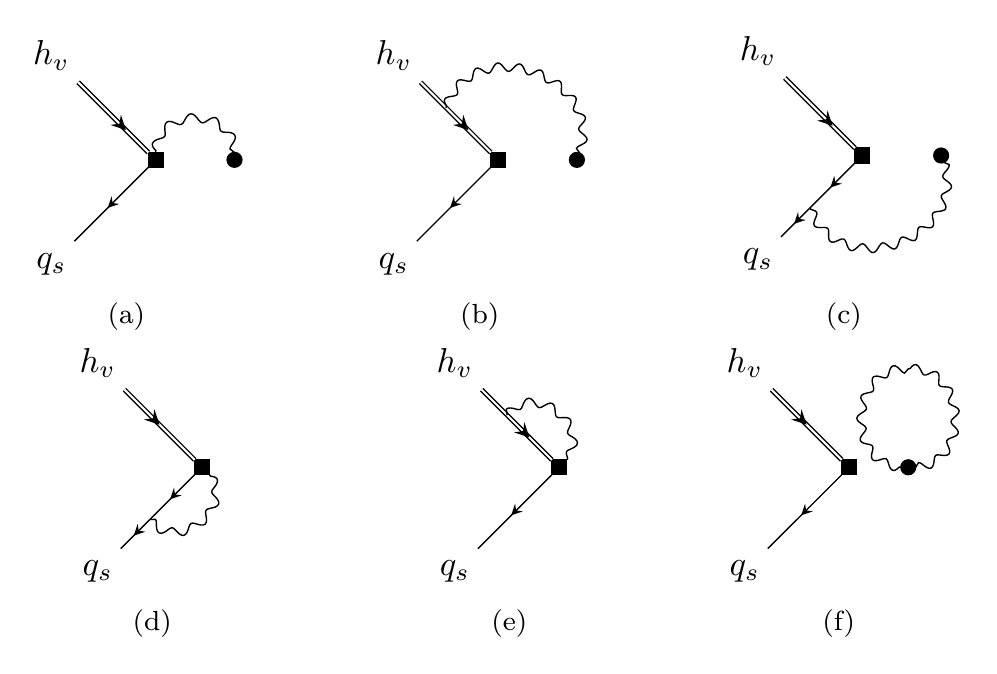}
	\caption{$\mathcal{O}(\alem)$ correction to the soft function. The square denotes the insertion of the finite-distance Wilson line $[tn_-,0]^{(q)}$ with charge $Q_q$ and the dot the insertion of the product of soft QED Wilson lines $S_{n_-}^{\dagger (Q_{M_1})}S_{n_+}^{\dagger (Q_{M_2})}$. The graph with a photon exchanged between the heavy-quark line and the spectator quark is UV-finite (in Feynman gauge) and is hence not shown.}
	\label{fig:BLCDAdia}
\end{figure}

In the following, we present details on the momentum-space calculation of the renormalization factor in \eqref{eq:B-Zfactor} including QED. We calculate the matrix element of the operator $\mathcal{O}_\otimes(\omega)$ on the left-hand side of \eqref{eq:BLCDAdef} with partonic external states $\langle 0|\mathcal{O}_{\qboth}(\omega)| \bar{q}_s(\omega') h_v\rangle$. The variable $\omega$ always refers to the light-cone momentum associated with the operator (i.e.~the Fourier conjugate to the position argument $t$), whereas the variable $\omega' \equiv n_- l$ is associated with the soft momentum $l$ of the external spectator-quark state. Since we restrict ourselves to the one-loop approximation, we identify $Z_\qboth$ and $\Gamma_\qboth$ with their one-loop expressions. We define
\begin{equation}
Z_\qboth(\omega, \omega',\mu) = 
\delta(\omega-\omega')+ 
\frac{\alpha_s}{4\pi} Z^{\rm{QCD}}(\omega, \omega',\mu)
+ \frac{\alem}{4\pi} Z^{\rm{QED}}_\qboth (\omega, \omega',\mu) \,.
\end{equation}
The $\mathcal{O}(\alem)$ contributions $Z^{\rm QED}_\qboth$ to the renormalization factor are represented by the diagrams in Fig.~\ref{fig:BLCDAdia}. Their computation differs in several aspects from the QCD case \cite{Lange:2003ff}. Most importantly, we need to introduce modified plus-distributions which mix positive momentum variable $\omega'>0$ to negative $\omega<0$ and vice versa. 
To highlight the difference of the QED generalization with respect to the standard $B$-meson LCDA, we explicitly compute the UV pole of diagram~a) in dimensional regularization (space-time dimension $d=4-2\epsilon$) using off-shell regularization for the infrared (IR) singularities of the partonic matrix element. The soft Wilson lines in the operator also inherit an off-shell infrared regulator from the hard-collinear propagators (see Appendix A of \cite{Beneke:2021pkl} for details). The Feynman rule for the finite-distance Wilson line $[tn_-,0]^{(q)}$ defined in \eqref{eq:finiteWL} for an outgoing photon with momentum $k$ and an incoming anti-quark $q_s$ with momentum $\ell$ (with $n_-l = \omega'$) reads
\begin{equation}
    \frac{Q_q e n_-^\mu}{n_-k + \delta_c}\left(\delta(\omega+n_-k -\omega') - \delta(\omega-\omega')\right) ,
\end{equation}
where $\delta_c = k_q^2/(n_+ k_q)+i0$. In diagram a), only the contraction of the finite-distance Wilson line with the soft Wilson line $S_{n_+}^{\dagger (Q_{M_2})}$ is UV divergent. This Wilson line contains the off-shellness $\delta_{\bar{c}}= k_q^2/(n_- k)+i0$. As the full anomalous dimension is independent of the IR regulator, we are free to choose ${\rm Re}(\delta_{c,\bar{c}})<0$. We label the charge of the spectator quark $q_s$ as $Q_{\rm sp}$ and, omitting the prefactor of $Q_{\rm sp} Q_{M_2}$, we obtain for diagram~a):
\begin{align} \label{eq:fig4a}
   &\mut^{2\eps} \int \frac{d^dk}{(2\pi)^d}  \frac{-2i}{k^2+i0}  \frac{e}{n_- k + \delta_c }  \frac{e}{n_+ k - \delta_{\bar{c}} }  \left(\delta(\omega+n_- k -\omega') - \delta(\omega-\omega')\right) \nonumber \\
     & =   \frac{-2 \alem}{(2\pi)^{d-2}}  \mut^{2\eps} \int_0^\infty \frac{d(n_-k)} {n_- k + \delta_c}   \int \frac{ d^{d-2}k_\perp}{-k_\perp^2 - \delta_{\bar{c}}n_-k}\left(\delta(\omega+n_- k -\omega') - \delta(\omega-\omega')\right) \nonumber \\
   & = \frac{-2 \alem}{(4\pi)^{1-\eps}}\left(\frac{\mut}{-\delta_{\bar{c}}}\right)^{\!\epsilon} \Gamma(\eps) \int_0^\infty \frac{d(n_- k)}{n_- k + \delta_c} \left( \frac{\mut}{n_- k} \right)^{\epsilon} \left(\delta(\omega+n_- k -\omega') - \delta(\omega-\omega')\right) \nn \\ 
    & = \frac{-2\alem}{(4\pi)^{1-\eps}} \left(\frac{\mut}{-\delta_{\bar{c}}}\right)^{\!\epsilon} \Gamma(\eps)\, \bigg\{ \frac{\theta(\omega'-\omega)}{\omega'-\omega +\delta_c} \left(\frac{\mut}{\omega'-\omega} \right)^{\eps} -\Gamma(\eps)\Gamma(1-\eps) \left(\frac{\mut}{\delta_c}\right)^{\epsilon} \delta(\omega-\omega') \bigg\} \,.
\end{align}
\vskip-0.1cm\noindent
Here we used $k^2 = (n_-k) (n_+k) + k_\perp^2$ in light-cone components. In the first line we integrated over $n_+k$ using the residue theorem which restricts $n_-k>0$. The result contains a local $\delta(\omega-\omega')$ and a non-local $\theta(\omega'-\omega)$ term. The latter arises from the first delta function $\delta(\omega +n_- k - \omega')$ due to the fact that $n_-k$ is positive. This term has a peculiar feature: it is non-zero for $\omega<\omega'$ and $\omega$ is not restricted to be positive. The support for negative $\omega$ is indeed required for consistency with the local limit $t\to 0$ of the left-hand side of \eqref{eq:BLCDAdef} before renormalization. This limit is obtained by integrating \eqref{eq:fig4a} over $\omega$, in which case the finite-distance Wilson line collapses to unity. Therefore diagram a) should vanish, which only happens when integrating $\omega$ over the entire real axis.
This means that even if we initially assume $\omega'>0$, negative $\omega$ will be generated by evolution. Therefore, different from QCD, the soft function acquires support for $-\infty <\omega <\infty$ for $Q_{M_2} \neq 0$ once electromagnetic effects are included. 

\begin{figure}[t]
\centering
\includegraphics[width=0.7\textwidth]{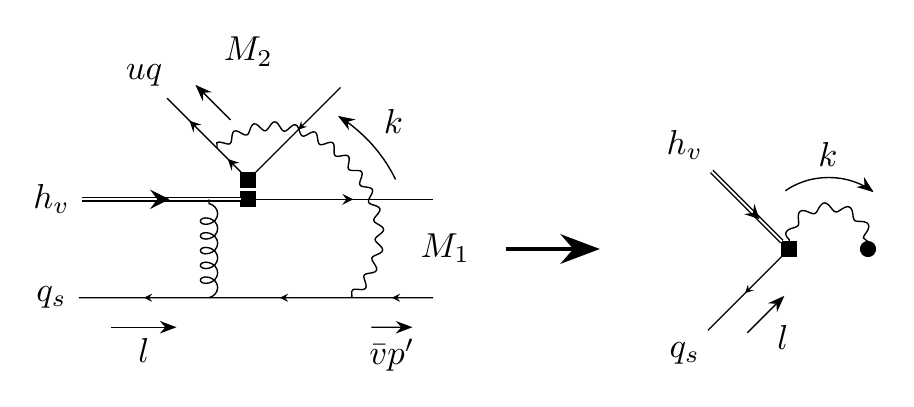}
	\caption{Example of a full theory (QCD$\times$QED) diagram giving rise to diagram a) of Fig.~\ref{fig:BLCDAdia} in the effective theory (HQET$\times$SCET$_{\rm II}$). In the full theory, a soft photon couples to the collinear and anti-collinear quarks of the two outgoing mesons, labeled by $M_1$ and $M_2$. The hard-collinear gluon (curly line) carries momentum $\ell-k-\bar{v}p'$, with $p'$ aligned to the light-like vector $n_-$. The hard-collinear gluon propagator at leading power in the effective theory is proportional to the inverse of $\omega = n_- (\ell - k) = \omega' - n_- k$, resulting in the standard $1/\omega$ inverse moment. Since the light-cone momentum of the soft photon (wavy line) can be of order $n_- k \sim \mathcal{O}(u n_- \cdot q) \sim \mathcal{O}(m_b \to \infty)$  we conclude that even for $\omega' > 0$ a contribution with $\omega = \omega'-n_-k<0$ will be generated.}
	\label{fig:BLCDAnegsupport}
\end{figure}

The support for $\omega<0$ in QED can also be understood by physics arguments. We recall that in QCD-only, in the $m_b\to \infty$ limit, the static HQET field serves as an infinite source of light-like momentum in the $n_+$ and $n_-$ direction, such that the spectator quark light-cone component $\omega'= n_-\ell$ can become infinitely large and $\omega$ extends to $\infty$. In QED, on the other hand, once $M_2$ is charged, the spectator quark $q_s$ can couple to the energetic quarks of the \emph{outgoing} meson $M_2$ (with momentum $q$) through the exchange of soft photons. These quarks are anti-collinear and therefore have large momentum components $n_-q \sim \mathcal{O}(m_b)$ in the $n_+$ direction. In the limit $m_b\to \infty$, the photons can thus carry away an infinite amount of light-cone momentum from the spectator quark, which is depicted in Fig.~\ref{fig:BLCDAnegsupport}. In this case, $\omega$ must extend to $-\infty$, which explains the negative support of the soft function. Therefore, we conclude that for outgoing, charged $M_2$ these functions have to be clearly distinguished from the standard QCD $B$-meson LCDA. 

The result \eqref{eq:fig4a} is part of the renormalization factor $Z^{\rm QED}_\qboth$ and has to be understood as a distribution that is integrated against a test function $\phi(\omega')$. It can also be understood as an operator valued distribution in $\omega$ rather than $\omega'$, which is discussed in Appendix B. To extract the UV anomalous dimension of the soft functions, we rewrite the non-local part of \eqref{eq:fig4a} as a plus-distribution for the separate cases $i)$ $\omega>0$ and $ii)$ $\omega<0$. In this way, we systematically expand the diagrammatic result in the off-shellnesses $\delta_{c,\bar{c}}$, which then cancel in the sum of all diagrams.

$i)$ First assuming $\omega>0$, we rewrite the non-local term 
with the help of the standard plus-distribution
\begin{align} \label{eq:PDwprime}
 \int_{-\infty}^\infty d\omega' \big[ \dots \big]_+ \phi(\omega') &= \int_{-\infty}^\infty d\omega' \big[ \dots \big] (\phi(\omega')-\phi(\omega))
\end{align}
into 
\begin{align}
\label{eq:exppd}
    &\theta(\omega) \int_{-\infty}^\infty d\omega' \frac{\theta(\omega'-\omega)}{\omega'-\omega+\delta_c} \left( \frac{\mut}{\omega'-\omega} \right)^{\!\eps} \phi(\omega') \nn \\  &= \theta(\omega) \int_{-\infty}^\infty d\omega' \omega' \left[\frac{\theta(\omega'-\omega)}{\omega'(\omega'-\omega+\delta_c)} \right]_+  \phi(\omega') +   \theta(\omega)\phi(\omega) \int_{\omega}^\infty d\omega'\frac{\omega }{\omega'(\omega'-\omega+\delta_c)} + \mathcal{O}(\eps) \nn \\
    &= \theta(\omega) \int_{-\infty}^\infty d\omega' \omega' \left[\frac{\theta(\omega'-\omega)}{\omega'(\omega'-\omega)} \right]_+  \phi(\omega') - \theta(\omega) \phi(\omega) \ln \frac{\delta_c}{\omega} + \mathcal{O}(\eps,\delta_c) \; .
\end{align}
We expanded the above result in $\eps$ under the assumption that the test function behaves like $1/\omega'$ up to logarithmic corrections. We therefore do not get any UV divergence in the $\omega'$ integration.\footnote{This behaviour is expected as the soft function should be considered part of a factorization formula in which it is convoluted with a collinear (jet) function.} In passing to the last line of \eqref{eq:exppd}, we further expanded the regulator $\delta_c$ in the plus-distribution, since it is IR finite for $\delta_c \to 0$.

$ii)$ For $\omega<0$, the non-local term $\theta(\omega'-\omega)/(\omega'-\omega)$ in \eqref{eq:fig4a} needs to be regulated with a distribution for $\omega'<0$, but not for $\omega'>0$. We therefore define modified plus-distributions (similar to the $+a$ distribution defined in \cite{Actis:2008rb})
\begin{align}  \label{eq:PDwprimeoplusminus}
 \int_{-\infty}^\infty d\omega' \big[ \dots \big]_{\oplus / \ominus} \,\phi(\omega') &=  \int_{-\infty}^\infty d\omega' \big[ \dots \big] (\phi(\omega')-\theta(\pm \omega') \phi(\omega)) \; .
\end{align}
The $\ominus$-distribution is needed in \eqref{eq:fig4a} as it regulates the $1/(\omega'-\omega)$ pole for $\omega<0$ and $\omega'<0$. The $\oplus$-distribution regulates the integration for $\omega>0$ and $\omega'>0$ and is needed for diagram~e) discussed below. Using the $\ominus$-distribution, we obtain
\begin{align}
   &\theta(-\omega) \int_{-\infty}^\infty d\omega' \frac{\theta(\omega'-\omega)}{\omega'-\omega+\delta_c} \left( \frac{\mut}{\omega'-\omega} \right)^{\!\eps} \phi(\omega') \nn \\  
   &= \theta(-\omega) \int_{-\infty}^\infty d\omega' \left[ \frac{\theta(\omega'-\omega)}{\omega'-\omega+\delta_c} \right]_\ominus \phi(\omega') + \theta(-\omega) \phi(\omega) \int_\omega^0 d\omega' \frac{1}{\omega'-\omega+\delta_c} + \mathcal{O}(\eps) \nn \\ 
   &= \theta(-\omega) \int_{-\infty}^\infty d\omega' \left[ \frac{\theta(\omega'-\omega)}{\omega'-\omega} \right]_\ominus \phi(\omega') - \theta(-\omega) \phi(\omega) \ln \frac{\delta_c}{-\omega} + \mathcal{O}(\eps,\delta_c) \ .
\end{align}

Finally, adding both contributions with the appropriate $\theta(\pm \omega)$ factors in \eqref{eq:fig4a}, the off-shell regulator $\delta_c$ cancels in the local terms  (but not $\delta_{\bar c}$) and the UV-divergent part of diagram a) becomes 
\begin{align} \label{eq:ZQEDa}
   Z^{\rm{QED},(a)}_\qboth  &=  Q_{\rm sp} Q_{M_2} \bigg\{\frac{2}{\epsilon} \theta(\omega) \omega' \left[ \frac{\theta(\omega'-\omega)}{\omega'(\omega'-\omega)} \right]_+ +\frac{2}{\epsilon} \theta(-\omega) \left[ \frac{\theta(\omega'-\omega)}{\omega'-\omega} \right]_{\ominus}  \nonumber \\
    &-  \left(\frac{2}{\eps^2} +\frac{2}{\epsilon} \ln \frac{\mu}{-\delta_{\bar{c}}} + \frac{2}{\epsilon} \theta(\omega) \ln \frac{\mu}{\omega} + \frac{2}{\epsilon} \theta(-\omega) \ln \frac{\mu}{-\omega}  \right) \delta(\omega-\omega') \bigg\} .
   \end{align}

The other diagrams are computed in a similar manner. Diagram b) can be straight-forwardly computed as it only contributes to the local part of the renormalization factor:
\begin{eqnarray}\label{eq:ZQEDb}
     Z^{\rm{QED},(b)}_\qboth  &=&   Q_d Q_{M_1} \left( \frac{1}{\eps^2} +\frac{2}{\epsilon} \ln \frac{\mu}{-\delta_c} \right)\delta(\omega-\omega') + Q_d Q_{M_2} \left(\frac{1}{\eps^2} +\frac{2}{\epsilon} \ln \frac{\mu}{-\delta_{\bar{c}}} \right)\delta(\omega-\omega')  \;.\nonumber\\[-0.3cm]  
\end{eqnarray}
\vskip-0.2cm\noindent
Diagrams c) and d) have again non-local terms that are proportional to 
\begin{align}
    \int_0^{\omega'} d (n_- k) f(n_-k) \delta(\omega+n_-k -\omega') = f(\omega'-\omega) \left[\theta(\omega) \theta(\omega'-\omega) - \theta(-\omega) \theta(\omega-\omega') \right] \; .
\end{align}
\vskip-0.05cm\noindent
This integral arises through the light-like separation of the spectator quark and yields separate contributions to positive and negative $\omega$. The non-local terms of these diagrams by themselves do not generate support for $\omega<0$ if $\omega'>0$, i.e. they do not mix negative and positive support. Therefore, the term proportional to $\theta(-\omega)$ is not present in QCD-only due to the purely positive supported $B$-LCDA. However, in QCD$\times$QED, the $\omega<0$ part has to be considered as it will be generated by diagram a). The complete one-loop results are 
\begin{align}
    Z^{\rm{QED},(c)}_\qboth  &=  \frac{2}{\epsilon} Q_{\rm sp} Q_{M_1} \bigg\{ -\theta(\omega)  \omega \left[\frac{ \theta(\omega'-\omega)}{\omega'(\omega'-\omega)}  \right]_+   - \theta(-\omega)  \omega\left[ \frac{ \theta(\omega-\omega')}{\omega'(\omega-\omega')} \right]_+  \nonumber \\ &+ \ln \left( \frac{\mu}{\omega-i0} \right)    \delta(\omega-\omega') - \ln\left(\frac{\mu}{-\delta_c}\right)
    \delta(\omega-\omega') \bigg\}\; , \\ 
    Z^{\rm{QED},(d)}_\qboth &= \frac{2}{\epsilon} Q_{\rm sp}^2 \bigg\{- \theta(\omega) \omega \left[ \frac{\theta(\omega'-\omega)}{\omega'(\omega'-\omega)}  \right]_+ -  \theta(-\omega)\omega \left[ \frac{\theta(\omega-\omega')}{\omega'(\omega-\omega')} \right]_+  - \delta(\omega-\omega') \bigg\} \; .  \label{eq:ZQEDcd}
\end{align}
We note that diagram c) arises from a similar full theory diagram as in Fig.~\ref{fig:BLCDAnegsupport}, but with the difference that the soft photon couples to the spectator quark on the left side of the hard-collinear interaction vertex. Therefore also this diagram mixes positive into negative support. Even though this cannot be seen from the UV-divergent part, this mixing appears manifestly in the finite terms. For diagram e) the situation is similar to diagram a), but now for $\omega>0$ in terms of the $\oplus$-distribution defined in \eqref{eq:PDwprimeoplusminus}. The result reads: 
\begin{align}
       Z^{\rm{QED},(e)}_\qboth &= Q_d Q_{\rm sp} \bigg\{-\frac{2}{\epsilon} \theta(\omega) \left[ \frac{\theta(\omega-\omega')}{\omega-\omega'} \right]_{\oplus} -\frac{2}{\epsilon} \theta(-\omega) \omega' \left[ \frac{\theta(\omega-\omega')}{\omega'(\omega-\omega')}  \right]_+ \nonumber  \\
    &+\left( \frac{1}{\eps^2} + \frac{2}{\epsilon} \theta(\omega) \ln \frac{\mu}{\omega} +\frac{2}{\epsilon} \theta(-\omega) \ln \frac{\mu}{-\omega}  \right)\delta(\omega-\omega')  \bigg\} \; . \label{eq:ZQEDe}
    \end{align}
Lastly, the tadpole diagram f) connects the two soft Wilson lines and is thus only non-zero if both final-state mesons are charged:
\begin{align}
    Z^{\rm{QED},(f)}_\qboth &= Q_{M_1} Q_{M_2} \left( -\frac{2}{\eps^2} -\frac{2}{\epsilon} \ln \frac{\mu}{-\delta_{\bar{c}}} -\frac{2}{\epsilon} \ln \frac{\mu}{-\delta_c} + \frac{2i\pi}{\epsilon}  \right) \delta(\omega-\omega') \; . 
    \label{eq:res}
\end{align}   
The off-shell regulators $\delta_c$ and $\delta_{\bar{c}}$ in $Z_{\otimes}^{\rm QED}$ cancel after adding the soft rearrangement factors $R_c$ and $R_{\bar{c}}$ via \eqref{eq:BLCDAdef}. These factors are defined via the absolute value of the Wilson line product in \eqref{eq:softsubtraction}, such that no spurious imaginary parts are introduced in the collinear functions of the process, as discussed in~\cite{Beneke:2021pkl}. Independent of this choice, the soft functions in QED are in general complex-valued functions, since they contain the soft rescattering phases of the two-body decay process.  

\subsection{Anomalous dimension for $\Phi_{B,\qboth}$}
\label{sec:anomdim}

Above we have shown how the (generalized) plus-distributions arise in the QCD$\times$QED soft function. It is convenient to define the following linear combinations of plus-distributions that appear in $Z^{\rm QCD}$ and $Z^{\rm QED}$:
\begin{align}\begin{aligned}\label{eq:FGfun}
    {F}^>(\omega,\omega') &= \omega \left[ \frac{\theta(\omega'-\omega)}{\omega'(\omega'-\omega)} \right]_+ + \left[ \frac{\theta(\omega-\omega')}{\omega-\omega'}\right]_{\oplus} \; , \\   
    {F}^<(\omega,\omega') &=  \omega \left[ \frac{\theta(\omega-\omega')}{\omega'(\omega-\omega')} \right]_++  \left[\frac{\theta(\omega'-\omega)}{\omega'-\omega} \right]_{\ominus} \; , \\
    {G}^>(\omega,\omega') &= \left(\omega+\omega'\right) \left[ \frac{\theta(\omega'-\omega)}{\omega'(\omega'-\omega)} \right]_+ -i\pi\delta(\omega-\omega') \; , \\
    {G}^<(\omega,\omega') &= \left(\omega+\omega'\right) \left[ \frac{\theta(\omega-\omega')}{\omega'(\omega-\omega')} \right]_+ +i\pi \delta(\omega-\omega') \; .
    \end{aligned}
\end{align}
The superscript $> (<)$ refers to $\omega>0$ ($\omega<0$). The generalized $\oplus$ and $\ominus$ distributions, as defined in \eqref{eq:PDwprimeoplusminus}, which appear in $F^>$ and $F^<$, give rise to mixing from $\omega'>0$ to $\omega<0$ and vice versa. We introduced a local imaginary part in the $G$-distributions, motivated by the results in the next section, where we consider inverse moments. 
The distributions in \eqref{eq:FGfun} in fact 
only appear in the linear combinations 
\begin{align}\begin{aligned}\label{eq:Hdef}
    H_\pm(\omega, \omega') \equiv \theta(\pm\omega) F^{> (<)} (\omega,\omega')+ \theta(\mp\omega) G^{< (>)}(\omega,\omega')  \; ,
    \end{aligned}
\end{align}
where the upper (lower) sign corresponds to $> (<)$. 
Combining the results for diagrams a)--f) of Fig. \ref{fig:BLCDAdia}, together with the QED heavy- and light-quark field-renormalization factors, we obtain
\begin{align}
\label{eq:ZQEDfnal}
    Z^{\rm QED}_\qboth (\omega,\omega') &= \bigg[ (Q_{\rm sp}^2 +2Q_{\rm sp} Q_{M_1}) \left( \frac{1}{\eps^2} + \frac{2}{\eps} \ln \frac{\mu}{\omega-i0} \right) \nn \\ & +\frac{2}{\eps} \left( i\pi (Q_{\rm sp} + Q_{M_1}) Q_{M_2} - \frac34 Q_{\rm sp}^2 - \frac12 Q_d^2  \right)  \bigg] \delta(\omega-\omega') \nn \\ 
    & - \frac{2}{\eps} Q_{\rm sp} \bigg[ Q_d H_+(\omega,\omega') - Q_{M_2}  H_-(\omega,\omega') \bigg] \; ,
\end{align} 
employing charge conservation $Q_d -Q_{\rm sp} = Q_{M_1}+ Q_{M_2}$ and $\ln\,(\mu/[\omega-i0]) = \ln\,(\mu/[-\omega]) + i\pi$ for $\omega<0$. As required, the final result is independent of the off-shell regulators $\delta_c$ and~$\delta_{\bar{c}}$. 

The QCD-only result can obtained from \eqref{eq:ZQEDfnal} by putting $Q_{M_1}\to0, Q_{M_2}\to 0$ and $Q_{\rm sp}^2, Q_d^2\to C_F$: 
\begin{align}
    Z^{\rm QCD}(\omega,\omega') &= C_F \bigg\{ \left( \frac{1}{\eps^2} + \frac{2}{\epsilon} \ln \frac{\mu}{\omega-i0} -\frac{5}{2\eps} \right)\delta(\omega-\omega') -\frac{2}{\epsilon}H_+(\omega,\omega') \bigg\} \; .
\end{align}
This expression differs from the standard QCD result \textit{in absence} of QED as it is also defined for $\omega<0$. This has an important consequence for the generalized QCD$\times$QED evolution kernel: the negative support of the soft function, once generated by QED effects, generates feedback from QCD corrections. The standard QCD result is obtained when integrating with a positively supported test function, in which case $H_+\to F$, where for $\omega$ and $\omega'>0$
\begin{equation}
    F(\omega,\omega') = \omega\left[ \frac{ ~\theta(\omega'-\omega)}{\omega'(\omega'-\omega)}\right]_+ + \left[\frac{\theta(\omega-\omega')}{\omega-\omega'}\right]_+  
    \label{eq:FLNdef}
\end{equation}
is the standard QCD distribution \cite{Lange:2003ff}. 

Finally, from  \eqref{eq:Gdef} and \eqref{eq:ZQEDfnal} we derive the anomalous dimension
\begin{align} \label{eq:BAD}
    \Gamma_\qboth(\omega,\omega')  &= \frac{\alpha_s C_F}{\pi} \bigg[ \left(\ln \frac{\mu}{\omega-i0} - \frac12 \right)\delta(\omega-\omega') -H_+(\omega,\omega') \bigg] \nn \\ &+ \frac{\alem}{\pi} \bigg[  \bigg(  (Q_{\rm sp}^2 +2Q_{\rm sp} Q_{M_1}) \ln \frac{\mu}{\omega-i0}  - \frac34 Q_{\rm sp}^2 -\frac12 Q_d^2  \\&+ i\pi (Q_{\rm sp} + Q_{M_1}) Q_{M_2} \bigg) \delta(\omega-\omega') - Q_{\rm sp} Q_d H_+(\omega,\omega') +  Q_{\rm sp} Q_{M_2} H_-(\omega,\omega') \bigg] \; , \nn 
\end{align}
for $\Phi_{B,\qboth}$,
where all distributions act on functions of the variable $\omega'$. The corresponding distributions acting on variable $\omega$ instead of $\omega'$ are given in Appendix~\ref{app:wdist}. 
In the following, we list the explicit results for the various charge combinations.

\subsubsection{\boldmath $\bar{B}^0 \to M_1^0 M_2^0 \; (\qboth = (0,0))$}

We first consider the case where both $M_1$ and $M_2$ are neutral, which fixes $Q_{\rm sp}=Q_d$. From the diagrams in Fig.~\ref{fig:BLCDAdia} only diagrams d) and e) contribute. For neutral $M_2$, the $\omega<0$ terms can be neglected because the appearing plus-distributions do not generate support for $\omega<0$ from an initial condition with $\omega'>0$. This implies that the distribution $F^>$ reduces to $F$ in \eqref{eq:FLNdef}. The result is a trivial extension of the QCD kernel: 
\begin{align}
\label{eq:G00}
    \Gamma_{00}(\omega,\omega') &=\frac{\alpha_s(\mu) C_F}{\pi} \bigg[\left(\ln \frac{\mu}{\omega} -\frac12 \right)\delta(\omega-\omega')  - {F}(\omega,\omega') \bigg] \nn \\ &+ \frac{\alem(\mu) Q_d^2}{\pi} \bigg[\left(\ln \frac{\mu}{\omega} -\frac54 \right)\delta(\omega-\omega')  - {F}(\omega,\omega') \bigg]\; .
    \end{align}
The different constants in the local part arise due to the subtraction of the QCD decay constant $F_{\rm stat}$ in~\eqref{eq:Gdef}.

\subsubsection{\boldmath $\bar{B}^- \to M_1^- M_2^0\; (\qboth = (-,0))$}

When $M_2$ is neutral, but $M_1$ charged and hence $Q_{\rm sp}=Q_u$, the diagrams in  Fig.~\ref{fig:BLCDAdia}a) and f) do not contribute. For $\qboth=(-,0)$, the definition of the soft function \eqref{eq:BLCDAdef} includes the factor $1/R_c$, which is necessary to cancel the off-shell regulator. Dropping the $\omega<0$ terms, for the reasons explained above, gives 
\begin{align}\label{eq:Gmin0}
    \Gamma_{-0}(\omega,\omega') &=  \frac{\alpha_s(\mu)  C_F}{\pi}  \bigg[\left(\ln \frac{\mu}{\omega} -\frac12 \right)\delta(\omega-\omega')  -{F}(\omega,\omega')  \bigg] \nn \\ &+\frac{\alem(\mu)}{\pi} \bigg[  \left( (Q_u^2 + 2Q_u Q_{M_1})\ln \frac{\mu}{\omega} - \frac{3}{4}Q_u^2 - \frac{1}{2}Q_d^2 \right) \delta(\omega-\omega') \nn \\ &- ( Q_u^2 + Q_u Q_{M_1} ) {F}(\omega,\omega') \bigg]  \, .
\end{align}
The result for $\Gamma_{00}$ is recovered by setting $Q_{M_1}=0$ and replacing the spectator quark charge $Q_u \to Q_d$. It will be relevant for the solution of the RGE that the coefficients multiplying the $\ln(\mu/\omega)$ term and the $-F$ distribution are not the same in QED, unlike in \eqref{eq:G00} and in QCD.

\subsubsection{\boldmath $\bar{B}^- \to M_1^0 M_2^- (\qboth = (0,-))$}
As soon as $M_2$ is charged, the situation is more involved since support for $\omega<0$ is generated. We add the QED contributions from \eqref{eq:res} for $Q_{M_1}=0$ and $Q_{\rm sp} = Q_u$, and split the anomalous dimension into its $\omega>0$ and $\omega<0$ part:
\begin{equation}
    \Gamma_\qboth (\omega,\omega') = \theta(\omega) \Gamma_\qboth^> (\omega,\omega') + \theta(-\omega) \Gamma_\qboth^< (\omega,\omega') \; .
\end{equation}
For $\omega<0$, we explicitly extract the imaginary part using  $\ln \mu/(\omega-i0) = \ln \mu/(-\omega) +i\pi$. The result takes the form 
\begin{align}
    \Gamma^>_{0-}(\omega,\omega') &= \frac{\alpha_s(\mu) C_F}{\pi}  \bigg[\left(\ln \frac{\mu}{\omega} -\frac12 \right)\delta(\omega-\omega')   -{F}^>(\omega,\omega')  \bigg] \nn \\ &+\frac{\alem(\mu)}{\pi} \bigg[ \left( Q_u^2 \ln \frac{\mu}{\omega}  - \frac{3}{4}Q_u^2 - \frac{1}{2}Q_d^2  \right) \delta(\omega-\omega')\nonumber \\ & - ( Q_u^2+Q_u Q_{M_2})  {F}^>(\omega,\omega') +  Q_u Q_{M_2}  \left( {G}^>(\omega,\omega') +i\pi \delta(\omega-\omega') \right)  \bigg]  \; , \\ 
    \Gamma^<_{0-}(\omega,\omega') &=  \frac{\alpha_s(\mu)  C_F}{\pi}  \bigg[\left(\ln \frac{\mu}{-\omega} -\frac12 \right)\delta(\omega-\omega')   -{G}^<(\omega,\omega') +i\pi \delta(\omega-\omega') \bigg] \nn \\ &+ \frac{\alem(\mu)}{\pi} \bigg[\left( Q_u^2\ln\frac{\mu}{-\omega}- \frac{3}{4}Q_u^2 - \frac{1}{2}Q_d^2  \right) \delta(\omega-\omega') \nn \\ &-(Q_u^2+ Q_u Q_{M_2}) \left( {G}^<(\omega,\omega')-i\pi \delta(\omega-\omega') \right)+Q_u Q_{M_2} {F}^<(\omega,\omega') \bigg] \; .
\end{align}
We observe that $\Gamma^<$ is obtained from $\Gamma^>$ by replacing $F^> \to G^<$ and $G^> \to F^<$. The explicit imaginary parts cancel those in $G^<$ and $G^>$. This implies that $\Gamma_{0-}$, and hence the soft function $\Phi_{0-}(\omega;\mu)$, are real-valued objects. 

\subsubsection{\boldmath $\bar{B}^0 \to M_1^+ M_2^- \; (\qboth = (+,-))$}
For the charge configuration $\qboth=(+,-)$, we set $Q_{\rm sp} = Q_d$, $Q_{M_1}=-Q_{M_2}$ and find that the off-shellness $\delta_c$ and $\delta_{\bar{c}}$ cancels among the diagrams in Fig.~\ref{fig:BLCDAdia}a)--e), while diagram f) is cancelled up to its imaginary part by the soft rearrangement factors $1/R_c$ and $1/R_{\bar{c}}$. The result is 
\begin{align}
    \Gamma^>_{+-}(\omega,\omega') &=\frac{\alpha_s(\mu)  C_F}{\pi}  \bigg[ \left(\ln \frac{\mu}{\omega} -\frac12 \right)\delta(\omega-\omega') -{F}^>(\omega,\omega')  \bigg]  \nn \\ &+\frac{\alem(\mu)}{\pi} \bigg[  \left( (Q_d^2 - 2 Q_d Q_{M_2} )\ln \frac{\mu}{\omega} + i\pi Q_{M_1} Q_{M_2} - \frac{5}{4}Q_d^2  \right)\delta(\omega-\omega')  \nn \\ & -Q_d^2 {F}^>(\omega,\omega')+ Q_d Q_{M_2}  \left( {G}^>(\omega,\omega') +i\pi \delta(\omega-\omega') \right) \bigg]  \; , \\
    \Gamma^<_{+-}(\omega,\omega') &=  \frac{\alpha_s(\mu)  C_F}{\pi}  \bigg[\left(\ln \frac{\mu}{-\omega} -\frac12 \right)\delta(\omega-\omega') -{G}^<(\omega,\omega') +i \pi \delta(\omega-\omega')   \bigg]  \nn \\ &+ \frac{\alem(\mu)}{\pi} \bigg[  \left(  (Q_d^2 - 2Q_d Q_{M_2} )\ln \frac{\mu}{-\omega} + i\pi (Q_{M_1}-Q_d) Q_{M_2} -\frac{5}{4}Q_d^2 \right) \delta(\omega-\omega') \nn \\ & -Q_d^2 \left( {G}^<(\omega,\omega')-i\pi \delta(\omega-\omega') \right) +Q_d Q_{M_2} {F}^<(\omega,\omega') 
     \bigg]  \; .  \label{eq:G+-}
\end{align}
We note that setting $Q_{M_2}=0$ eliminates the distribution $F^<$, which mixes positive into negative support. We then recover the case \eqref{eq:G00} and \eqref{eq:Gmin0} where the soft function only acquires support for $\omega>0$. Moreover, setting $Q_{M_2}\to Q_\ell$ gives the anomalous dimension for the soft function in the process $B_q\to \mu^+\mu^-$. The soft function for $B$-meson decay into 
two electrically charged mesons is complex, as expected, due to 
soft final-state rescattering. The imaginary part already appears 
in the anomalous dimension.

\subsection{First inverse moment of $\Phi_{B,\qboth}$}
\label{sec:momentRGEs}

Besides the soft functions $\Phi_{B,\qboth}$ themselves, the first inverse moment and its logarithmic modification are of particular interest, because they appear at leading power in the factorization of exclusive $B$ decays. In factorization theorems, the function $\Phi_{B,\qboth}$ will always be convoluted with a (hard-)collinear function. It is therefore natural to focus on the evolution of these moments rather than the evolution of the soft functions themselves. For hard-exclusive processes the collinear function is proportional to $1/(\omega-i0)$, where the $i0$-prescription is inherited from a (hard-)collinear propagator. Loop corrections introduce logarithmic corrections, which retain the same $i0$-prescription due to the analytic structure of the collinear function. The definition of the first inverse moments is analogous to QCD with the exception that the $i0$-prescription must be kept, since the integration in $\omega$ extends over the entire real axis. Hence, we define for $n>0$
\begin{align} \label{eq:momentdef}
    \frac{1}{\lambda_{B}(\mu)} &= \int_{-\infty}^\infty \frac{d\omega}{\omega-i0} \,\Phi_{B,\qboth}(\omega;\mu) \nn \; , \\  \sigma_{n}(\mu) &=\lambda_B(\mu) \int_{-\infty}^\infty \frac{d\omega}{\omega-i0} \ln^n \left( \frac{\mut}{\omega-i0}\right) \Phi_{B,\qboth}(\omega;\mu)  \; ,
\end{align}
where the logarithmic moments are normalized to $\lambda_B^{-1}(\mu)$ and the scale $\mut$ is chosen to be a fixed reference scale, so that the evolution in $\mu$ is only from $\Phi_{B,\qboth}(\omega;\mu)$. 

In QCD, the leading-twist $B$-meson LCDA $\phi^+_B(\omega)$ can 
be shown to behave as $\phi^+_B(\omega)\propto \omega$ as 
$\omega\to 0$ at sufficiently high scales $\mu$, and the above moments, defined on positive 
$\omega$, are well-defined. For the cases when $Q_{M_2}=0$, which also have positive support, the moments remain well-defined with QED included, see Appendix \ref{sec:asymptoticQM2}. However, when $Q_{M_2}\not=0$, the function 
$\Phi_{B,\qboth}(\omega;\mu)$ will acquire a non-zero value or logarithmically diverge at 
$\omega=0$ through evolution, so that the $i0$-prescription in 
the denominator must be kept. Nevertheless, the inverse-logarithmic moments 
defined in \eqref{eq:momentdef} are finite, but higher inverse moments such as $1/(\omega-i0)^2$ do not 
exist, similar as in QCD, as will be shown in the following section.
The integration in \eqref{eq:momentdef} with the $i0$-prescription can in principle generate an imaginary part, which 
provides a new source of rescattering phases in hard spectator-interactions. In QCD, these can only arise from hard or hard-collinear loops.

For the moments in \eqref{eq:momentdef}, we can derive a coupled system of the RGEs from \eqref{eq:BAD} by computing the right-hand side of 
\begin{align}
   \frac{d}{d\ln \mu} \left( \frac{\sigma_n(\mu)}{\lambda_B(\mu)}\right) = -\int_{-\infty}^\infty \frac{d\omega}{\omega-i0} \ln^n \left( \frac{\mut}{\omega-i0}\right) \int_{-\infty}^\infty d\omega'~\Gamma_\qboth(\omega,\omega';\mu) \Phi_{B,\qboth}(\omega';\mu) \; . 
\end{align}
After exchanging the order of integrations and performing the $\omega$-integral first, the diagonal terms proportional to $\delta(\omega-\omega')$ are evaluated trivially and the calculation reduces to the evaluation of the plus-distributions $H_\pm$ in the kernel. For this, we need of these distributions to act in the variable $\omega$, which are given in \eqref{eq:FGinomega}, and denoted with a superscript $\omega$. We define $H_\pm^{(\omega)}$ analogous to \eqref{eq:Hdef}. For the $H_+^{(\omega)}$ distribution, we find the same result as in QCD
\begin{align} \label{eq:logF}
        \int_{-\infty}^\infty d\omega \frac{\ln^n \frac{\mut}{\omega-i0}}{\omega-i0}  \; H_+^{(\omega)}(\omega,\omega')  = \frac{2n!}{\omega'-i0}  \sum_{k=1}^{\lfloor n/2 \rfloor}\frac{\zeta_{2k+1}}{(n-2k)!} \ln^{n-2k} \frac{\mut}{\omega'-i0} \;  , 
\end{align}
where $\zeta$ is the Riemann zeta function and $n\geq 0$. The evaluation of the distribution $G^>_\omega$ in this calculation is subtle and requires a careful treatment of the $i0$-prescription to keep track of all imaginary parts. We find that \eqref{eq:logF} is real, which motivated including the explicit $i\pi\delta(\omega-\omega')$ term in the definition \eqref{eq:FGfun}. The agreement with the QCD result can be understood from the fact that $H_+^{(\omega)}$ reduces to $F_\omega$ on functions with positive support. Evaluating \eqref{eq:logF} with $H_-^{(\omega)}$ on the left-hand side produces the same result. Hence, both distributions agree on the inverse-logarithmic moments,
\begin{align} \label{eq:logdiffFG}
 \int_{-\infty}^\infty d\omega \, \frac{\ln^n \frac{\mut}{\omega-i0}}{\omega-i0} \; \bigg( H_+^{(\omega)}(\omega,\omega') - H_-^{(\omega)}(\omega,\omega') \bigg) = 0 \; .
\end{align}
This implies that the analogous integral in \eqref{eq:logdiffFG} over $F_\omega^>(\omega,\omega')-G_\omega^>(\omega,\omega')$ and $G_\omega^<(\omega,\omega')-F_\omega^<(\omega,\omega')$ vanishes for $\omega'>0$ and $\omega'<0$, respectively. The surprising consequence of this result is that in \eqref{eq:BAD} we can now replace the distribution $H_-$, which causes the mixing from positive to negative values of $\omega$, by $H_+$. We emphasize that \eqref{eq:logdiffFG} only holds on this particular function space, namely the inverse-logarithmic moment space, and is in general not true, for example, for the generalized moments mentioned at the end of Section~\ref{sec:invmomentsolution}.

Using \eqref{eq:logF} and \eqref{eq:logdiffFG}, we can derive the RGE for the first inverse and inverse-logarithmic moments. For $n=0$, the $\omega$ integral over the plus-distributions vanishes and we obtain 
\begin{align} \label{eq:firstmomRGE}
    \frac{d}{d\ln \mu} \lambda_B^{-1}(\mu) =& \frac{\alpha_s C_F}{\pi} \bigg[-\sigma_1(\mu) + \ln \frac{\mut}{\mu} + \frac12 \bigg]  \lambda_B^{-1}(\mu) \nn \\
    & + \frac{\alem}{\pi} \bigg[ (Q_{\rm sp}^2+2 Q_{\rm sp} Q_{M_1})\left( -\sigma_1(\mu) + \ln \frac{\mut}{\mu} \right) + \frac34 Q_{\rm sp}^2 + \frac12 Q_d^2 \nn \\ &-i\pi (Q_{\rm sp}+Q_{M_1}) Q_{M_2} \bigg]  \lambda_B^{-1}(\mu)  \; . 
\end{align}
For the logarithmic moments, we find
\begin{align} \label{eq:sigmaRGE}
    \frac{d\sigma_n}{d\ln \mu} =& \frac{\alpha_s C_F}{\pi} \bigg[ -\sigma_{n+1} + \sigma_n \sigma_1 + 2n! \sum_{k=1}^{\lfloor n/2 \rfloor}\frac{\zeta_{2k+1}}{(n-2k)!} \sigma_{n-2k} \bigg] \nn \\&+ \frac{\alem}{\pi} \bigg[(Q_{\rm sp}^2 +2 Q_{\rm sp} Q_{M_1} )(-\sigma_{n+1} + \sigma_n \sigma_1)  \nonumber \\
    & + (Q_{\rm sp}^2 +Q_{\rm sp} Q_{M_1} )\, 2n! \sum_{k=1}^{\lfloor n/2 \rfloor}\frac{\zeta_{2k+1}}{(n-2k)!} \sigma_{n-2k}  \bigg]\; .
\end{align}
For $\alem= 0$, this agrees with the QCD result \cite{Bell:2008er}.\footnote{Our QCD result differs by a term $n\sigma_{n-1}$ compared to \cite{Bell:2008er}, since we distinguish the reference scale $\tilde{\mu}$ from the renormalization scale $\mu$.} The form of the RGEs is very similar to QCD with the exception that the relative charge factors of the logarithmic moments in the square brackets are different. We note that under evolution from $\mu_0$ to $\mu$ the inverse moment $\lambda_B^{-1}$  acquires the complex phase 
$\exp\,(-i\pi (Q_{\rm sp} + Q_{M_1}) Q_{M_2}\, \frac{\alem}{\pi}\ln\frac{\mu}{\mu_0})$, when the scale dependence of the electromagnetic coupling is neglected. The logarithmic moments $\sigma_n(\mu)$ on the other hand remain real under scale evolution as long as all $\sigma_n(\mu_0)$ are real at some scale $\mu_0$, since they only mix into themselves and there are no imaginary terms in their RGEs.

\section{Analytic solution to the evolution equation}
\label{sec:analytic}

In this section, we solve the RGE \eqref{eq:B-RGE} using the QED-generalized anomalous dimension \eqref{eq:BAD} for the soft functions $\Phi_{B,\qboth}$. In addition, we derive solutions to the evolution equations \eqref{eq:firstmomRGE} and \eqref{eq:sigmaRGE} for the first inverse and inverse-logarithmic moments. We further discuss the asymptotic behaviour of the soft functions and the existence of their moments. For practical applications, we expand and solve the RGE of the soft functions to first-order in the electromagnetic coupling, since QED evolution effects are small at scales below $m_b$.

\subsection{Evolution equation in Laplace space}
In QCD, the evolution equation is typically solved by performing a Laplace or Mellin transformation, so that the RGE becomes local in the first argument of the LCDA and can be solved in the corresponding space \cite{Bell:2013tfa,Braun:2014owa}. In the following, we will use these techniques to derive analytical expressions for the evolved function $\Phi_{B,\qboth}$ and its inverse moments, given a general initial condition. To this end, we divide the support of the soft function into
\begin{align} \label{eq:supportdiff}
    \Phi_{B}(\omega;\mu) = \theta(\omega) \Phi_>(\omega;\mu) + \theta(-\omega) \Phi_< (\omega;\mu) \; ,
\end{align}
where we drop the charge label $\qboth$ in this section. We define the Laplace transform with respect to the variable $\ln(\mu/\omega)$ separately for $\omega>0$ by
\begin{align}
    \label{eq:mellintrafo}
    \tilde{\Phi}_>(\eta;\mu) &= \int_0^\infty \frac{d\omega}{\omega} \left(\frac{\mu}{\omega}\right)^{\!\eta} \Phi_>(\omega;\mu) \; , \nn\\
    \Phi_>(\omega;\mu) &= \int_{c-i\infty}^{c+i\infty} \frac{d\eta}{2\pi i} \left( \frac{\mu}{\omega}\right)^{\! -\eta} \tilde{\Phi}_>(\eta;\mu) \; ,
\end{align}
and for $\omega<0$ by
\begin{align}\label{eq:meltra}
    \tilde{\Phi}_<(\eta;\mu) &= \int^{\infty}_0 \frac{d\omega}{\omega} \left( \frac{\mu}{\omega} \right)^{\!\eta} \Phi_<(-\omega;\mu) \; \nn\\
    \Phi_<(-\omega;\mu)& = \int_{c-i\infty}^{c+i\infty} \frac{d\eta}{2\pi i} \left(\frac{\mu}{\omega} \right)^{\!-\eta} \tilde{\Phi}_<(\eta;\mu) \; .
\end{align}
Both transformations are related linearly to the function
\begin{align} \label{eq:generalinvint}
    \tilde{\Phi}_{B} (\eta;\mu) = \int_{-\infty}^\infty \frac{d\omega}{\omega-i0} \left( \frac{\mu}{\omega-i0} \right)^{\!\eta} \Phi_{B}(\omega;\mu) = 
    \tilde{\Phi}_{>}(\eta;\mu) - e^{i\pi \eta} \tilde{\Phi}_{<}(\eta;\mu)  \; .
\end{align}
We emphasize that the function $\tilde{\Phi}_B$ is {\em defined} by the right-hand side of \eqref{eq:generalinvint}, which means that it does not have the interpretation of an integral transformation on its own. In fact, there is no inverse transformation to obtain $\Phi_B$ back from $\tilde{\Phi}_B$. However, taking the limit $\eta \to 0$, the function $\tilde{\Phi}_B$ reproduces the inverse-logarithmic moments, see \eqref{eq:momentsfromphi} and the discussion below.

We can use the results from Appendix \ref{app:sec:purepowers} for the (generalized) plus-distributions acting on pure powers to derive the RGE for the above functions in Laplace space. We obtain a coupled system for $\tilde{\Phi}_>$ and $\tilde{\Phi}_<$ for $-1 < {\rm Re}(\eta) < 0$ given by
\begin{align}
    \left( \frac{d}{d\ln \mu} -\eta \right) \tilde{\Phi}_>(\eta;\mu) & = \frac{\alpha_s C_F}{\pi} \bigg[ -H_\eta -H_{-\eta} -\partial_\eta + \frac12 \bigg]\tilde{\Phi}_>(\eta;\mu)  \nn \\&+ \frac{\alpha_s C_F}{\pi} \Gamma(-\eta) \Gamma(1+\eta) \tilde{\Phi}_<(\eta;\mu) \nn \\
    &+ \frac{\alem}{\pi} \bigg[ -(Q_{\rm sp}^2 +2Q_{\rm sp} Q_{M_1}) \partial_\eta  + \frac34 Q_{\rm sp}^2 + \frac12 Q_d^2 -i\pi Q_{M_1} Q_{M_2} \nn \\&- Q_{\rm sp} Q_d (H_\eta + H_{-\eta}) + Q_{\rm sp} Q_{M_2} (H_{-\eta} + H_{-1-\eta}) \bigg] \tilde{\Phi}_>(\eta;\mu) \nn \\&+ \frac{\alem}{\pi} Q_{\rm sp} Q_d \Gamma(-\eta) \Gamma(1+\eta) \tilde{\Phi}_<(\eta;\mu) \; . \label{eq:phigRGE} \\
    \left( \frac{d}{d\ln \mu} -\eta \right) \tilde{\Phi}_<(\eta;\mu) &= \frac{\alpha_s C_F}{\pi} \bigg[ - H_{-\eta}-H_{-1-\eta} -\partial_\eta + \frac12 \bigg]\tilde{\Phi}_<(\eta;\mu) \nn \\&+\frac{\alem}{\pi} \bigg[ -(Q_{\rm sp}^2 +2Q_{\rm sp} Q_{M_1}) \partial_\eta   + \frac34 Q_{\rm sp}^2 + \frac12 Q_d^2 -i\pi (Q_{\rm sp} + Q_{M_2}) Q_{M_1}  \nn \\&- Q_{\rm sp} Q_d (H_{-\eta} + H_{-1-\eta}) + Q_{\rm sp} Q_{M_2} (H_\eta + H_{-\eta} )  \bigg] \tilde{\Phi}_<(\eta;\mu) \nn \\ &- \frac{\alem}{\pi} Q_{\rm sp} Q_{M_2} \Gamma(-\eta) \Gamma(1+ \eta)  \tilde{\Phi}_>(\eta;\mu) \; ,
\end{align}
where $H_\eta = \gamma_E+ \psi(\eta+1)$ is the Harmonic number function, which is related to the digamma function $\psi(\eta)= \Gamma'(\eta)/\Gamma(\eta)$. We can use these equations to derive the RGE for $\tilde{\Phi}_B$ and find
\begin{align}
\label{eq:mellinRGE}
\left( \frac{d}{d \ln \mu} - \eta \right) \tilde{\Phi}_{B}(\eta;\mu) &= \frac{\alpha_s C_F}{\pi} \bigg[ -H_\eta-H_{-\eta}-\partial_\eta + \frac12 \bigg] \tilde{\Phi}_{B}(\eta;\mu) \nonumber \\ &+ \frac{\alem}{\pi} \bigg[ Q_{\rm sp} Q_{M_1} (H_\eta + H_{-\eta}) - (Q_{\rm sp}^2 +2Q_{\rm sp} Q_{M_1}) \left(  H_\eta + H_{-\eta} + \partial_\eta \right)    \nn \\ &+ \frac{3}{4} Q_{\rm sp}^2 + \frac{1}{2} Q_d^2 - i\pi (Q_{\rm sp} + Q_{M_1} )Q_{M_2} \bigg]\tilde{\Phi}_{B}(\eta;\mu) \; ,
\end{align}
where we used the identity $\Gamma(-\eta)\Gamma(1+\eta) = e^{i\pi \eta}(i\pi + H_\eta - H_{-1-\eta})$ and its implication for the combination $H_{-1+\eta} + H_{-1-\eta} = H_\eta + H_{-\eta}$.
In \eqref{eq:logdiffFG} we observed that the distributions $H_\pm$ agree on the function space of inverse-logarithmic moments. This statement can be extended to pure powers with the corresponding $i0$-prescriptions as in \eqref{eq:generalinvint}, see App. \ref{app:sec:purepowers}. This implies that \eqref{eq:mellinRGE} can also be derived without the prior separation into $\omega>0$ and $\omega<0$ pieces, and holds for $-1<{\rm Re}(\eta)<1$, as can be seen from the analytic structure.

To illustrate the strategy of the solution method, we recall some properties of the solution  $\phi_B^+$ in QCD, which obeys the Laplace transform \eqref{eq:mellintrafo}. In this case, the RGE is
\begin{align} \label{eq:QCDRGE}
    \frac{d}{d\ln \mu} \phi_B^+(\omega;\mu) = - \int_{0}^\infty d\omega'~\Gamma^{\rm QCD}(\omega,\omega';\mu)~\phi_B^+(\omega';\mu) \; .
\end{align}
The QCD anomalous dimension is defined by $\Gamma^{\rm QCD} = \Gamma_\qboth |_{\alem = 0}$ in \eqref{eq:BAD}, which reduces to the standard QCD evolution kernel \cite{Lange:2003ff} since we assume $\phi_B^+(\omega;\mu)$ to vanish for $\omega<0$. If functions with negative support are included, the extra terms in $\Gamma^{\rm QCD}$ contribute and the lower bound of the integration extends to $-\infty$. The Laplace transform converges for $-1<{\rm Re}(\eta)<1$ and yields
\begin{align} \label{eq:QCDRGElaplace}
    \left( \frac{d}{d\ln \mu} -\eta \right) \tilde{\phi}_B^+(\eta;\mu) = \frac{\alpha_s C_F}{\pi} \left[-H_\eta -H_{-\eta}-\partial_\eta  +\frac12 \right] \tilde{\phi}_B^+(\eta;\mu) \; .
\end{align}
 The solution of \eqref{eq:QCDRGElaplace} can be derived analytically. We define 
\begin{align} \label{eq:VaQCD}
    V^{\rm QCD}(\mu,\mu_0)  =&-\int_{\mu_0}^\mu \frac{d\mu'}{\mu'} \frac{\alpha_s(\mu') C_F }{\pi} \bigg[  \ln \frac{\mu'}{\mu_0} -\frac12  \bigg] \; , \nonumber \\ 
    a^{\rm QCD}(\mu,\mu_0) = &-\int_{\mu_0}^\mu \frac{d\mu'}{\mu'} \frac{\alpha_s(\mu')C_F}{\pi} \; .
\end{align}
The dependence of these variables on $\mu$ and $\mu_0$ will be implicitly understood if not indicated otherwise. In addition, for readability, we drop the superscript QCD from here until \eqref{eq:QCDMeijerG} below. The general solution, including the inverse transformation for \eqref{eq:QCDRGE}, is given by \cite{Bell:2013tfa} 
\begin{align} 
    \tilde{\phi}_B^+(\eta;\mu) &= e^{V+2\gamma_E a} \left( \frac{\mu}{\mu_0} \right)^{\!\!\eta} \frac{\Gamma(1-\eta) \Gamma(1+\eta+a)}{\Gamma(1+\eta)\Gamma(1-\eta-a)} \,\tilde{\phi}_B^+(\eta+a;\mu_0) \; , \label{eq:expanalyticlap} \\
    \phi^+_B(\omega;\mu) &= e^{V+2\gamma_E a} \int_0^\infty \frac{d\omega'}{\omega'} \left( \frac{\mu_0}{\omega'} \right)^{\! a} G_a \!\left( \frac{\omega}{\omega'}\right) \phi^+_B (\omega',\mu_0) \; . \label{eq:expanalytic}
\end{align}
The parameter of the inverse transformation \eqref{eq:mellintrafo} lies in the convergence strip $-1<c<1$. Here and in the following, we present the solution in terms of a convolution of the initial condition with the Meijer-G function 
\begin{align} \label{eq:QCDMeijerG}
    G_a\!\left( \frac{\omega}{\omega'} \right) &\equiv G^{1,1}_{2,2} \bigg( {{-a\ , 1-a } \atop {1 \, ,0}}   \bigg|~ \frac{\omega}{\omega'}  \bigg) \\
    &= \frac{\Gamma(2+a)}{\Gamma(-a)} \left(\frac{\omega'}{\max(\omega,\omega')} \right)^a \frac{\min(\omega,\omega')}{\max(\omega,\omega')}~_2F_1\left(1+a,2+a;2;\frac{\min(\omega,\omega')}{\max(\omega,\omega')}\right) \; . \nn 
\end{align}
The detailed definition and some properties of this type of functions are given in Appendix \ref{app:MeijerG}. Most notably, $G_a(z)$ is singular for $z\to 1$, but integrable. The representation in \eqref{eq:QCDMeijerG}, although rather unconventional, enables us to write the results in a compact form. 

\subsection{Solution for the first inverse (logarithmic) moments} 
\label{sec:invmomentsolution}

We defined the inverse-logarithmic moments in \eqref{eq:momentdef} and derived their RGEs in Sec. \ref{sec:momentRGEs}. The result is an infinite-dimensional coupled system of equations given by \eqref{eq:firstmomRGE} and \eqref{eq:sigmaRGE}. In the following, we derive a solution to these equations. This is done by deriving the solution for the RGE \eqref{eq:mellinRGE} and then computing the first inverse-logarithmic moments from \eqref{eq:generalinvint} via 
\begin{align} \label{eq:momentsfromphi}
    \lambda_B^{-1}(\mu) &=  \lim_{\eta \to 0} \tilde{\Phi}_B(\eta;\mu) = \lim_{\eta \to 0} \big\{  \tilde{\Phi}_{>}(\eta;\mu) - e^{i\pi \eta} \tilde{\Phi}_{<}(\eta;\mu) \big\} \ ,\\
    \sigma_n(\mu) &=   \lim_{\eta \to 0}  \lambda_B(\mu) \bigg( \partial_\eta - \ln \frac{\mu}{\mut} \bigg)^{\!n} \tilde{\Phi}_{B}(\eta;\mu) \label{eq:momentsfromphisig} \; .
\end{align}
Note that the second inverse moment is proportional to $1/(\omega-i0)^2$ and therefore corresponds to the limit $\eta \to 1$.

For the solution $\tilde{\Phi}_B$, we define the QED-generalized evolution variables 
\begin{eqnarray} 
\label{eq:VQED}
    V &=& V^{\rm QCD} -\int_{\mu_0}^\mu \frac{d\mu'}{\mu'} \frac{\alem(\mu')}{\pi} \,\bigg[ (Q_{\rm sp}^2 +2Q_{\rm sp} Q_{M_1}) \ln \frac{\mu'}{\mu_0} - \frac34 Q_{\rm sp}^2 - \frac12  Q_d^2 
\nonumber\\  
&& + \,i\pi (Q_{\rm sp} +Q_{M_1} ) Q_{M_2} \bigg] \; , \nonumber \\ 
    a &=& a^{\rm QCD} -\int_{\mu_0}^\mu \frac{d\mu'}{\mu'} \frac{\alem(\mu') }{\pi} (Q_{\rm sp}^2 +2Q_{\rm sp} Q_{M_1}) \; ,
\end{eqnarray}
where $V^{\rm QCD}$ and $a^{\rm QCD}$ are defined in \eqref{eq:VaQCD}. Again, the dependence on $\mu$ and $\mu_0$ in the argument of $V$ and $a$ is implicitly understood. Furthermore, we define evolution functions
\begin{align}
\label{eq:calF}
    \mathcal{F}(\eta;\mu,\mu_0) &= \exp \bigg\{ \int_{\mu_0}^{\mu} \frac{d\mu'}{\mu'}  \frac{\alem(\mu')Q_{\rm sp} Q_{M_1}}{\pi} (H_{\eta+a(\mu,\mu')} + H_{-\eta-a(\mu,\mu')}) \bigg\}\; , \\
    \mathcal{G}(\eta;\mu,\mu_0) &= \exp \bigg\{ \int_{\mu_0}^{\mu} \frac{d\mu'}{\mu'}  \frac{\alem(\mu')Q_{\rm sp} Q_{M_1}}{\pi} (H_{-\eta-a(\mu,\mu')} + H_{-1-\eta-a(\mu,\mu')}) \bigg\}\; .
\end{align}
Adapting \cite{Beneke:2021pkl,Liu:2020eqe}, the general solution is
\begin{align} \label{eq:mellinsol}
\tilde{\Phi}_{B}(\eta;\mu) &= e^{\Vqed +2\gamma_E \aqed} \left( \frac{\mu}{\mu_0}\right)^{\!\eta} \frac{\Gamma(1-\eta)}{\Gamma(1+\eta)}\frac{\Gamma(1+\eta+\aqed)}{\Gamma(1-\eta-\aqed)}\mathcal{F}(\eta;\mu,\mu_0) \tilde{\Phi}_{B}(\eta+\aqed;\mu_0) \; .
\end{align}
We can use this solution to calculate the moments from \eqref{eq:momentsfromphisig}. For $\eta=0$, we find the first inverse moment
\begin{align} \label{eq:lambdasol}
    \lambda_B^{-1}(\mu) &= e^{\Vqed+2\gamma_E \aqed} \frac{\Gamma(1+\aqed)}{\Gamma(1-\aqed)} \mathcal{F}(0;\mu,\mu_0) \int_{-\infty}^\infty \frac{d\omega}{\omega-i0} \left( \frac{\mu_0}{\omega-i0} \right)^{\!\aqed} \Phi_{B}(\omega;\mu_0)  \; .
\end{align} 
For physically relevant scales, we have $-1<a<0$. The existence of the $\omega$ integral is naturally related to the asymptotic behaviour of the function $\Phi_B$, which is discussed in Sec. \ref{sec:allorder}. The solution for $\sigma_1$ can be either computed from \eqref{eq:momentsfromphisig} and \eqref{eq:mellinsol} or derived from the RGE \eqref{eq:firstmomRGE}. We obtain 
\begin{align}
    \sigma_1(\mu) &= H_\aqed + H_{-\aqed} + \ln \frac{\mut}{\mu_0} + 
\frac{\displaystyle \int_{-\infty}^\infty \frac{d\omega}{\omega-i0} \left(\frac{\mu_0}{\omega-i0} \right)^{\!\aqed} \ln \frac{\mu_0}{\omega-i0} \,\Phi_{B}(\omega;\mu_0) }{\displaystyle  \int_{-\infty}^\infty \frac{d\omega}{\omega-i0} \left( \frac{\mu_0}{\omega-i0} \right)^{\!\aqed} \Phi_{B}(\omega;\mu_0)} \nn \\ &+\int_{\mu_0}^{\mu} \frac{d\mu'}{\mu'}  \frac{\alem(\mu') Q_{\rm sp} Q_{M_1}}{\pi} (H'_{\aqed(\mu,\mu')} -H'_{-\aqed(\mu,\mu')}) \; . \label{eq:sigma1sol}
\end{align}
In this way, we can consecutively derive the solution for the infinite-dimensional coupled system of equations given by \eqref{eq:firstmomRGE} and \eqref{eq:sigmaRGE}. For $\alem=0$ and restricting the support of $\Phi_{B}$ to $\omega>0$, the expressions for $\lambda_B$ and $\sigma_1$ agree with the QCD results \cite{Bell:2008er}, up to a difference due to the reference scale $\mut \neq \mu$. We stress that \eqref{eq:lambdasol} and \eqref{eq:sigma1sol} correspond to the exact solution of \eqref{eq:firstmomRGE} and \eqref{eq:sigmaRGE} in QCD$\times$QED.

There is also an alternative way to arrive at the results \eqref{eq:lambdasol} and \eqref{eq:sigma1sol}, in which we could have considered $\Phi_{B}$ as an auxiliary function. For this, we assume that $H_+ = H_-$ holds already for the RGE of $\Phi_{B}$. This identification eliminates the distributions that mix positive to negative support. Hence, we can consider a reduced RGE only for $\omega>0$ 
\begin{align} \label{eq:red1}
    \frac{d}{d\ln \mu} \Phi^{\rm red}_{B}(\omega;\mu) = - \int_0^\infty d\omega'~\Gamma^{\rm red}_{\qboth}(\omega,\omega';\mu)\ \Phi^{\rm red}_{B} (\omega';\mu) \; , 
\end{align} 
where the anomalous dimension is given in terms of the standard QCD distribution $F(\omega,\omega')$, defined in \eqref{eq:FLNdef}, by
\begin{align} \label{eq:red2}
    \Gamma_{\qboth}^{\rm red}(\omega,\omega') &= \frac{\alpha_s C_F}{\pi} \bigg[ \bigg(\ln \frac{\mu}{\omega} -\frac12 \bigg)\delta(\omega-\omega')  - F(\omega,\omega') \bigg] + \frac{\alem}{\pi} \bigg[ \bigg( (Q_{\rm sp}^2 + 2Q_{\rm sp} Q_{M_1})\ln \frac{\mu}{\omega}  \nn \\ &- \frac34 Q_{\rm sp}^2 - \frac12 Q_d^2 + i\pi(Q_{\rm sp} + Q_{M_1})Q_{M_2} \bigg)\delta(\omega-\omega') -(Q_{\rm sp}^2+Q_{\rm sp} Q_{M_1})F(\omega,\omega') \bigg] \; .
\end{align}
\vskip-0.1cm\noindent
We obtain \eqref{eq:lambdasol} and \eqref{eq:sigma1sol} using the same steps as before, but with the function $\Phi^{\rm red}_{B,\qboth}$. Since $\Phi^{\rm red}_{B,\qboth}$ has no physical interpretation, one should rather express the appearing integrals in terms of the inverse-logarithmic moments at the initial scale $\mu_0$. For arbitrary $n$ and a function with positive support, this can be done with
\begin{align}
    \int_0^\infty \frac{d\omega}{\omega} \left(\frac{\mu_0}{\omega} \right)^a \ln^n \frac{\mu_0}{\omega} \,\Phi_{B,\qboth}(\omega;\mu_0) = \lambda_B^{-1}(\mu_0) \left(\frac{\mu_0}{\mut} \right)^a \sum_{k=0}^n \left( n\atop k \right) \ln^{n-k} \frac{\mu_0}{\mut} \sum_{l=0}^\infty \frac{a^l \sigma_{k+l}(\mu_0)}{l!} \,.
\end{align}
\vskip-0.1cm\noindent
An analogous equation holds for functions with negative support, so that we obtain the same result for the solution in terms of $\lambda_B(\mu_0)$ and $\sigma_n(\mu_0)$ in both cases.

Finally, we remark that the discussion for the inverse-logarithmic moments is not valid for more general moments, as appear for instance in $B\to \gamma^* \ell \nu$ \cite{Beneke:2021rjf,Wang:2021yrr}, where the moment $1/(\omega-n_-q-i0)$ is shifted by the momentum component $n_-q$ of the virtual photon. In this case, \eqref{eq:logdiffFG} no longer holds and the corresponding RGE does not reduce to a simple form as in \eqref{eq:mellinRGE} or \eqref{eq:red1} and \eqref{eq:red2} since such moments cannot be derived from $\tilde{\Phi}_B$, given in \eqref{eq:mellinsol}, alone. Hence, the solution for these moments needs to be derived from a generalized combination of the all-order solution for $\tilde{\Phi}_>$ and $\tilde{\Phi}_<$.


\subsection{All-order solution for $\Phi_B$} 
\label{sec:allorder}

The all-order solution $\Phi_B$ to the RGE \eqref{eq:B-RGE} can be formally constructed from the inverse transformations of $\tilde{\Phi}_>$ and $\tilde{\Phi}_<$ in \eqref{eq:mellintrafo} and \eqref{eq:meltra}. To this end, we want to find the complete solution in Laplace space. In the last section we derived the solution for $\tilde{\Phi}_B$ in \eqref{eq:mellinsol}. We can use \eqref{eq:generalinvint} to write  
\begin{align} \label{eq:phisrecover}
    \tilde{\Phi}_<(\eta;\mu) = e^{-i\pi \eta} (\tilde{\Phi}_>(\eta;\mu) - \tilde{\Phi}_{B,\qboth}(\eta;\mu) ) \; .
\end{align}
The RGE \eqref{eq:phigRGE} for $\tilde{\Phi}_>$ then turns into
\begin{align}
     \left( \frac{d}{d\ln \mu} -\eta \right) \tilde{\Phi}_>(\eta;\mu) & = \frac{\alpha_s C_F}{\pi} \bigg[ -H_{-\eta} -H_{-1-\eta} -\partial_\eta +i\pi + \frac12 \bigg]\tilde{\Phi}_>(\eta;\mu)  \nn  \\
    &+ \frac{\alem}{\pi} \bigg[ (Q_{\rm sp}^2 +2Q_{\rm sp} Q_{M_1}) (-H_{-\eta} -H_{-1-\eta} -\partial_\eta) \nn  \\
    & + Q_{\rm sp} Q_{M_1}(H_{-\eta}+H_{-1-\eta}) + \frac34 Q_{\rm sp}^2 + \frac12 Q_d^2 -i\pi (Q_{\rm sp}+ Q_{M_1}) Q_{M_2}  \nn  \\
    &+i\pi (Q_{\rm sp} Q_d + Q_{\rm sp} Q_{M_2})  \bigg] \tilde{\Phi}_>(\eta;\mu) \nn \\&- \frac{\alpha_s C_F + \alem Q_{\rm sp} Q_d}{\pi}  e^{-i\pi \eta} \Gamma(-\eta) \Gamma(1+\eta) \tilde{\Phi}_{B,\qboth}(\eta;\mu) \; .
\end{align}
Hence, we need to solve again a first-order differential equation for $\tilde{\Phi}_>$ but with $\tilde{\Phi}_B$ treated as an inhomogeneous part. This equation can by solved by the standard variation of constants methods. We define 
\begin{align}
    \hat{a}(\mu,\mu_0) = - \int_{\mu_0}^\mu \frac{d\mu'}{\mu'} \frac{\alpha_s(\mu') C_F + \alem(\mu')(Q_{\rm sp} Q_d + Q_{\rm sp} Q_{M_2})}{\pi} 
\end{align}
and find 
\begin{align} \label{eq:phiallorder}
    \tilde{\Phi}_>(\eta;\mu) &= e^{V+2\gamma_E a} \bigg( \frac{\mu}{\mu_0} \bigg)^\eta \bigg[ \frac{\Gamma(-\eta)\Gamma(1-\eta)}{\Gamma(-\eta-a)\Gamma(1-\eta-a)} e^{-i\pi\hat{a}} \mathcal{G}(\eta;\mu,\mu_0) \tilde{\Phi}_>(\eta+a;\mu_0)  \nn \\ 
    &- \frac{\Gamma(-\eta)\Gamma(1-\eta)\Gamma(1+\eta+a)}{\Gamma(1-\eta-a)} \mathcal{F}(\eta;\mu,\mu_0) e^{-i\pi \eta} \tilde{\Phi}_B(\eta+a;\mu_0)  \\
    &\times \int_{\mu_0}^\mu \frac{d\mut}{\mut} \frac{\alpha_s(\mut)C_F +\alem(\mut) Q_{\rm sp} Q_d}{\pi} e^{ -i\pi a(\mu,\mut) -i\pi \hat{a}(\mu,\mut)} (\mathcal{F}^{-1} \mathcal{G})(\eta;\mu,\mut)  \bigg] \; . \nn
\end{align}
Taking the limit $\alem \to 0$, we recover the QCD-only expression in \eqref{eq:expanalyticlap}. Note that we can simplify the combination 
\begin{align}
(\mathcal{F}^{-1} \mathcal{G})(\eta;\mu,\mut) = \exp \bigg\{  \int_{\mut}^{\mu} \frac{d\mu'}{\mu'} \alem(\mu') Q_{\rm sp} Q_{M_1} \cot \pi(\eta+a(\mu,\mu')) \bigg\} \; ,
\end{align}
by using the identity $H_{-1-\eta} - H_\eta = \pi \cot \pi \eta$. We refrain from giving the explicit expression for $\tilde{\Phi}_<$, which can be trivially obtained from \eqref{eq:mellinsol}, \eqref{eq:phisrecover} and \eqref{eq:phiallorder}. In principle, we can compute the function $\Phi_B(\omega;\mu)$ with numerical methods from the inverse Laplace transform. However, to estimate the QED corrections which are expected to be numerically small it will be sufficient to consider the solution up to $\mathcal{O}(\alem)$ and use the analytic expressions derived in Sec. \ref{sec:firstorder}.

We can use the all-order solution \eqref{eq:phiallorder} to study the behaviour of the soft function $\Phi_B$ in momentum space. There are two cases of particular interest. $i)$ For $\omega \to 0$, we want to investigate the analytic structure, especially whether the soft function can be regarded as an analytic function in the variable $\omega-i0$. $ii)$ For $\omega \to \pm \infty$, the power law behaviour is relevant for the convergence of the integration in \eqref{eq:momentdef}. For our purposes, we consider the exponential model $\Phi_B(\omega)=\omega/\omega_0^2\,e^{-\omega/\omega_0}\,\theta(\omega)$, which reads in Laplace space:
\begin{align}
    \tilde{\Phi}_>(\eta+a;\mu_0) = \frac{1}{\omega_0} \bigg( \frac{\mu_0}{\omega_0} \bigg)^{\eta+a} \Gamma(1-\eta-a)\; , \qquad \tilde{\Phi}_<(\eta+a;\mu_0) = 0\,.
\end{align}
The asymptotic behaviour is dictated by the analytic structure of the RGE in Laplace space, in particular by the poles of the Harmonic numbers and Gamma functions from the generalized distributions $F$ and $G$. In the following, we use the inverse transformation \eqref{eq:mellintrafo} to derive the asymptotic behaviour for the two cases above. In our derivation, we closely follow the analysis for the soft approximation of light mesons in \cite{Beneke:2021pkl}. In this approximation, one of the light-meson quarks becomes soft while the other one is nearly static, reproducing the situation for a heavy meson. Note that for realistic scales $\mu$ and $\mu_0$, we have $-1<a<0$.

We discuss case $i)$ first. For $\omega \to 0$ we can deform the contour $c\pm i\infty$ of the inverse transformation \eqref{eq:mellintrafo} to enclose all poles and branch cuts in the right half-plane with respect to ${\rm Re}(\eta)=c$, where $-1-a<c<0$. We extract the singular term for $\eta \to 0$ from $\Gamma(-\eta)$ and $\mathcal{G}(\eta)$ in \eqref{eq:phiallorder} and find the leading-power behaviour
\begin{align}
    \Phi_>(\omega\to 0;\mu) \sim \frac{1}{\omega_0} \int_C \frac{d\eta}{2\pi i} \bigg(\frac{\omega}{\omega_0}\bigg)^\eta \bigg( \frac{1}{-\eta} \bigg)^{1+p(\mu)} = \frac{1}{\omega_0} \frac{1}{ \Gamma(1+p(\mu))} \bigg( -\ln \frac{\omega}{\omega_0} \bigg)^{p(\mu)}  \; ,
\end{align}
where
\begin{align} \label{eq:pdef}
    p(\mu) = \frac{-\alem(\mu) Q_{\rm sp} Q_{M_1}}{\alpha_s(\mu) C_F +\alem(\mu) (Q_{\rm sp}^2 + 2 Q_{\rm sp} Q_{M_1})}
\end{align}
and the contour $C$ is chosen according to Appendix B.2 of \cite{Beneke:2021pkl}. From \eqref{eq:phisrecover}, we observe that $\tilde{\Phi}_<$ has a similar behaviour but with the corresponding $i0$-prescription ensured by the exponential prefactor which 
combines to $e^{-i\pi \eta} (-\omega)^\eta  = (\omega-i0)^\eta$ for $\omega<0$. Therefore, we can write the asymptotic expansion of the soft function for $\omega\to 0$ as
\begin{align} \label{eq:phizero}
    \Phi_B(\omega\to 0;\mu) &= \frac{1}{\omega_0} \frac{\kappa}{\Gamma(1+p(\mu))} \bigg( -\ln \frac{\omega-i0}{\omega_0} \bigg)^{p(\mu)} + \mathcal{O}\bigg(\frac{\omega}{\omega_0} \bigg)  \; ,
\end{align}
where $\kappa$ is a dimensionless constant. The difference between the two support regimes $\omega>0$ and $\omega<0$ is given by the function $\tilde{\Phi}_B$ in \eqref{eq:mellinsol} whose branch cut starts at ${\rm Re}(\eta)=1$ and extends to the right. This cut can only give rise to $\mathcal{O}(\omega/\omega_0)$ corrections in \eqref{eq:phizero}, so that the linear contributions do not need to retain the $i0$-prescription of the leading-power result. We remark that \eqref{eq:phizero} holds when negative support is included, which is, however, not always the case. For  $Q_{M_2}=0$, we only have positive support and the solution reduces to a much simpler result, see Appendix \ref{sec:asymptoticQM2}. We conclude that for $\qboth=(0,-)$, where $p=0$, the soft function acquires a constant value at $\omega=0$. However, for $\qboth=(+,-)$ we have $p>0$ due to the values of the quark charges and the soft function logarithmically diverges. These observations agree with our findings from the numerical solution in Fig. \ref{fig:0minus}. 

For case $ii)$ $\omega\to\pm \infty$ we deform the integration contour to enclose the poles and cuts on the left half-plane with respect to ${\rm Re}(\eta)=c$. The relevant contribution enters from the singular contribution $\eta\to -1-a$ and we obtain
\begin{align} \label{eq:phiinf}
    \Phi_B(\omega \to \pm \infty;\mu) \sim \frac{1}{\omega_0} \bigg(  \frac{\pm \omega}{\omega_0} \bigg)^{-1-a} \ln^{p(\mu_0)} \bigg(\frac{\pm \omega}{\omega_0} \bigg) \; .
\end{align}
The power-like behaviour is similar to QCD-only \cite{Lange:2003ff} and proportional $\omega^{-1-a}$, but with the QED-generalized evolution variable $a$ defined in \eqref{eq:VQED}. Note that, even though the particular asymptotic behaviour \eqref{eq:phizero} and \eqref{eq:phiinf} refers to the exponential model, the analysis can be easily applied to a more general class of models with similar conclusions, depending on their analytic structure in Laplace space.

\subsubsection{Comment on the existence of inverse moments} 
\label{sec:momex}
The asymptotic behaviour for $\omega \to 0$ and $\omega \to \pm \infty$ is relevant to understand the existence of different moments of the soft function. We recall that in QCD-only the LCDA vanishes linearly for $\omega \to 0$ and the RGE generates a behaviour of the form $\phi_B^+(\omega;\mu)\propto \omega^{-1-a^{\rm QCD}}$ for $\omega \to \infty$. Consequently, only the first inverse-logarithmic moments exist after scale evolution while all other moments are divergent. As one would expect and based on our computations, the same holds true in QCD$\times$QED. For $\omega\to 0$, we found the logarithmic dependence in \eqref{eq:phizero} and that the linear and higher order terms do not depend analytically on $\omega-i0$. Thus, for the inverse-logarithmic moments \eqref{eq:momentdef}, we can deform the contour away from $\omega=0$ for the leading terms, which would otherwise be singular, while the non-analytic $\mathcal{O}(\omega/\omega_0)$ terms in \eqref{eq:phizero} are already finite. This guarantees the existence of the first inverse-logarithmic moments. On the other hand, the second and higher inverse moments do not exist, since the non-analytic $\mathcal{O}(\omega/\omega_0)$ terms are still singular at $\omega=0$. In Laplace space, this is connected to the poles of $\Gamma(1-\eta)$ from $\tilde{\Phi}_B$ in \eqref{eq:mellinsol} for $\eta = 1,2,3,\dots$ On the other hand, the soft function falls off like $\omega^{-1-a}$ for $\omega \to \pm \infty$. Hence, all non-negative moments of $\Phi_B$ diverge since $-1<a<0$. Due to the behaviour for $\omega \to \pm \infty$, we further remark that the second inverse moment in QCD$\times$QED can be related to the derivative of the soft function using integration by parts
\begin{align}
    \int_{-\infty}^\infty \frac{d\omega}{(\omega-i0)^2} \Phi_{B}(\omega;\mu) =  \int_{-\infty}^\infty  \frac{d\omega}{\omega-i0} \partial_\omega \Phi_{B}(\omega;\mu) \; .
\end{align}
For the case of $\qboth=(0,-)$, in which the soft function is continous at $\omega=0$, this shows directly that the soft function cannot be differentiable at this point, because the above discussion shows that the left-hand side does not exist. In the general case, for a positively supported initial condition and $p(\mu)>0$, the log-divergence in \eqref{eq:phizero} is an $\mathcal{O}(\alem^2)$ effect, so that it does not appear in the numerical evaluation of the first-order solution in Sec. \ref{sec:numerics}. 

\subsection{Analytic solution for $\Phi_{B,\qboth}$ to $\mathcal{O}(\alem)$}  \label{sec:firstorder}

In practice, the QED corrections are much smaller than the 
QCD ones, therefore we consider only the first-order correction 
in $\alem$ while summing the QCD logarithms to all orders. The solution of the RGE to first-order in the electromagnetic coupling can be obtained from the expansion of the formal solution in \eqref{eq:phiallorder}. However, we obtain much simpler expressions right away by expanding and solving the RGE \eqref{eq:B-RGE} to $\mathcal{O}(\alem)$ as we do in the following. To this end, we expand the soft function as
\begin{align}
    \label{eq:alemexpansion}
    \Phi(\omega;\mu) = \phi_B^+(\omega;\mu) + \frac{\alem}{\pi} \Phi^{(1)} (\omega;\mu) + \mathcal{O}(\alem^2) \; ,
\end{align}
where $\phi_B^+(\omega;\mu)$ is the standard QCD solution with positive support only. All QED effects are therefore part of $\Phi^{(1)}(\omega;\mu) \sim \mathcal{O}(\ln^2( \mu/\mu_0),\alpha_s(\mu_0)/\alpha_s(\mu))$. The first-order QED solution can be obtained with the help of methods developed in \cite{Bosch:2003fc,Liu:2020eqe}. To derive an equation for $\Phi^{(1)}(\omega;\mu_0)$, we need to consistently expand the RGE to first order and therefore split the one-loop kernel into its QCD and QED contribution,
\begin{equation}
    \Gamma_\qboth (\omega,\omega',\mu) = \Gamma^{\rm QCD}(\omega, \omega', \mu) + \frac{\alem}{\pi} \Gamma^{\rm QED}_\qboth(\omega, \omega',\mu)  \; .
\end{equation}
Analogously to \eqref{eq:supportdiff}, we divide the QED correction into its positive and negative domain
\begin{equation}\label{eq:qedphi}
    \Phi^{(1)}(\omega;\mu) = \theta(\omega)  \Phi_{>}^{(1)}(\omega; \mu)  +\theta(-\omega)  \Phi_{<}^{(1)}(\omega; \mu) \;  .
\end{equation}
After subtracting \eqref{eq:QCDRGE}, the RGE at $\mathcal{O}(\alem)$ reads
\begin{align} \label{eq:firstorderRGE}
    \frac{d}{d\ln \mu} \Phi^{(1)}(\omega;\mu) =& - \int_{-\infty}^\infty d\omega'\;\Gamma^{\rm QCD}(\omega,\omega';\mu) \,\Phi^{(1)}(\omega';\mu) \nn \\
    & - \int_0^{\infty} d\omega' \; \Gamma^{\rm QED}_\qboth(\omega,\omega';\mu) \,\phi^+_B(\omega';\mu)  \; .
\end{align}
Both terms on the right-hand side are in general non-zero for $\omega<0$. The QED kernel in the second term produces negative support from $\omega'>0$ while the QCD kernel in the first term already contains $\omega<0$ contributions that need to be included due to the negative support of $\Phi^{(1)}$. Inserting \eqref{eq:qedphi} for $\Phi^{(1)}$ turns the RGE into a coupled system for $\Phi^{(1)}_>$ and $\Phi^{(1)}_<$. Using the definitions \eqref{eq:FGfun} and \eqref{eq:FLNdef}, the explicit form of this system is
\begin{align}
    \frac{d}{d\ln \mu}  \Phi^{(1)}_>(\omega;\mu) &= \frac{\alpha_s C_F}{\pi} \int_0^\infty d\omega' \left[ \left(-\ln \frac{\mu}{\omega} +\frac12 \right)\delta(\omega-\omega') + F(\omega,\omega') \right] \Phi^{(1)}_>(\omega';\mu) \nn \\ &+ \frac{\alpha_s C_F}{\pi} \int_{-\infty}^0 d\omega' F^>(\omega,\omega') \Phi^{(1)}_<(\omega';\mu) + \int_0^\infty d\omega' \bigg[(Q_{\rm sp}^2 +Q_{\rm sp} Q_{M_1} ) F(\omega,\omega')\nn \\&+  Q_{\rm sp} Q_{M_2}(F(\omega,\omega')-G^>(\omega,\omega')) -\bigg( (Q_{\rm sp}^2 +2Q_{\rm sp} Q_{M_1})\ln \frac{\mu}{\omega}\nn \\&+i\pi (Q_{\rm sp} +Q_{M_1}) Q_{M_2} -\frac34 Q_{\rm sp}^2 -\frac12 Q_d^2\bigg)\delta(\omega-\omega') \bigg]\phi^+_B(\omega';\mu) \; ,\label{eq:firstorderRGE1} \\ 
    \frac{d}{d\ln \mu} \Phi^{(1)}_<(\omega;\mu) &= \frac{\alpha_s C_F}{\pi} \int_{-\infty}^0 d\omega'\left[ \left(-\ln \frac{\mu}{-\omega}-i\pi +\frac12 \right)\delta(\omega-\omega')  + G^<(\omega,\omega') \right]\Phi^{(1)}_<(\omega';\mu) \nn \\ 
    &- Q_{\rm sp} Q_{M_2} \int_0^\infty d\omega' F^<(\omega,\omega')\phi^+_B(\omega';\mu)\; . \label{eq:firstorderRGE2}
\end{align}
The first equation holds for $\omega>0$, the second for $\omega<0$. The crucial observation is that the RGE for $\Phi_<^{(1)}$ is independent of $\Phi^{(1)}_>$, hence it can be solved on its own using the Laplace transformation. The solution for $\Phi_<^{(1)}$ can be then be inserted into \eqref{eq:firstorderRGE1}. In Laplace space, we obtain
\begin{align}
    \left( \frac{d}{d\ln \mu} - \eta \right) \tilde{\Phi}^{(1)}_>(\eta;\mu) &= \frac{\alpha_s C_F}{\pi} \left[-H_{\eta} -H_{-\eta} -\partial_\eta +\frac12 \right] \tilde{\Phi}^{(1)}_>(\eta;\mu)   \nn \\
    &-\frac{\alpha_s C_F}{\pi} \Gamma(\eta) \Gamma(1-\eta) \tilde{\Phi}^{(1)}_<(\eta;\mu)  + \bigg[ Q_{\rm sp} Q_{M_1} (H_\eta + H_{-\eta}) \nn \\ &+Q_{\rm sp} Q_{M_2} (H_{-1-\eta}-H_\eta)  - (Q_{\rm sp}^2 + 2Q_{\rm sp} Q_{M_1}) (H_\eta +H_{-\eta} +\partial_\eta)  \nn \\ &-i\pi Q_{M_1} Q_{M_2} + \frac34 Q_{\rm sp}^2 + \frac12 Q_d^2 \bigg] \tilde{\phi}_B^+(\eta;\mu)   \; , \label{eq:begMellin} \\ 
    \left( \frac{d}{d\ln \mu} - \eta \right) \tilde{\Phi}^{(1)}_<(\eta;\mu) &= \frac{\alpha_s C_F}{\pi} \left[-H_{-\eta} -H_{-1-\eta} -\partial_\eta +\frac12 \right] \tilde{\Phi}^{(1)}_<(\eta;\mu) \nn \\ 
    &+ Q_{\rm sp} Q_{M_2} \Gamma(\eta) \Gamma(1-\eta) \tilde{\phi}_B^+(\eta;\mu) \label{eq:negMellin} \; , 
\end{align}
where $-1< {\rm Re}(\eta) < 0$ as discussed in Appendix  \ref{app:sec:purepowers}. This implies that even though we seemingly perform two independent transformations of $\Phi_>^{(1)}$ and $\Phi_<^{(1)}$, we need to choose the region $-1 < c < 0$ for both transformations. This region corresponds to the strip of convergence, for which the integrals of the generalized plus-distributions in \eqref{eq:FGfun} with pure powers exist. A complete list of these distributions acting on pure powers can be found in Appendix~\ref{app:sec:purepowers}. 

The general solution to \eqref{eq:negMellin} can be found (similarly to \cite{Liu:2020eqe}) by a variation of constants and is given by
\begin{align} \label{eq:QM2MellinsolNeg}
     \tilde{\Phi}_<^{(1)}(\eta;\mu) &= e^{V+2\gamma_E a} \left( \frac{\mu}{\mu_0} \right)^{\!\eta} \frac{\Gamma(1-\eta)\Gamma(-\eta)}{ \Gamma(1-\eta-a)\Gamma(-\eta-a)} \bigg[ \tilde{\Phi}_<^{(1)}(\eta+a;\mu_0) \nn \\ &- Q_{\rm sp} Q_{M_2} \Gamma(-\eta-a) \Gamma(1+\eta+a)  \tilde{\phi}_B^+(\eta+a;\mu_0) \ln \frac{\mu}{\mu_0} \bigg]  \; .
\end{align} 
Plugging this result into \eqref{eq:begMellin}, we obtain the solution for $\tilde{\Phi}_>^{(1)}$: \begin{align} \label{eq:posMellin}
    \tilde{\Phi}_>^{(1)}(\eta;\mu) &=  e^{V+2\gamma_E a} \left( \frac{\mu}{\mu_0} \right)^{\!\eta} \frac{\Gamma(1-\eta)\Gamma(1+\eta+a)}{\Gamma(1+\eta)\Gamma(1-\eta-a)} \Bigg[ \tilde{\Phi}^{(1)}_>(\eta+a;\mu_0) \nn \\ 
    &+ \int_{\mu_0}^\mu  \frac{d\mu'}{\mu'} \frac{\alpha_s(\mu') C_F}{\pi} \frac{\Gamma^2(-\eta-a(\mu,\mu')) \Gamma^2(1+\eta+a(\mu,\mu'))}{\Gamma(-\eta-a)\Gamma(1+\eta+a) } \bigg\{ \tilde{\Phi}^{(1)}_<(\eta+a;\mu_0) \nn \\
    &- Q_{\rm sp} Q_{M_2} \Gamma(-\eta-a) \Gamma(1+\eta+a) \tilde{\phi}_B^+(\eta+a;\mu_0) \ln \frac{\mu'}{\mu_0} \bigg\}  \nn \\ 
    &+ \bigg( \int_{\mu_0}^\mu  \frac{d\mu'}{\mu'}  \{ Q_{\rm sp} Q_{M_1}( H_{\eta+a(\mu,\mu')} + H_{-\eta-a(\mu,\mu')} ) \nn \\ 
    &+ Q_{\rm sp} Q_{M_2}(H_{-1-\eta-a(\mu,\mu')}-H_{\eta+a(\mu,\mu')} ) \}    \nn \\ &-(Q_{\rm sp}^2 +2 Q_{\rm sp} Q_{M_1} ) \left\{ \frac12 \ln^2 \frac{\mu}{\mu_0} + (H_{\eta+a} + H_{-\eta-a} + \partial_\eta) \ln \frac{\mu}{\mu_0}  \right\} \nn \\ &+ \left[ \frac34 Q_{\rm sp}^2 + \frac12 Q_d^2 - i\pi Q_{M_1} Q_{M_2} \right] \ln \frac{\mu}{\mu_0} \,\bigg) \,\tilde{\phi}_B^+(\eta+a;\mu_0) \Bigg] \; .
\end{align}
Note that we can evaluate the integral in the second line in terms of harmonic numbers when it is multiplied with $\tilde{\Phi}_<^{(1)}(\eta+a;\mu_0)$ in the curly bracket. At this point, we have to perform the inverse transformation to obtain the function $\Phi^{(1)}(\omega;\mu)$. To this end, we rewrite the harmonic numbers as 
\begin{equation}
    H_\eta = -\int_0^1 dx \left[\frac{1}{1-x} \right]_+ x^\eta = - \int_0^1 \frac{dx}{1-x} \,(x^\eta-1)  \; .
\end{equation}
Afterwards, we can perform some of the $\mu'$-integrals in \eqref{eq:posMellin}. We define the function 
\begin{equation}
    h(x) \equiv \int_{\mu_0}^\mu \frac{d\mu'}{\mu'} x^{a(\mu,\mu')} =\frac{2 \pi}{\beta_0^{\rm QCD}+ 2C_F \ln x } \left[ \frac{1}{\alpha_s(\mu)} - \frac{1}{\alpha_s(\mu_0)} \left( \frac{\alpha_s(\mu)}{\alpha_s(\mu_0) }\right)^{ \frac{2C_F}{\beta_0^{\rm QCD}} \ln x } \right]\; ,
\end{equation}
which evaluates to the last expression in the one-loop 
approximation. The inverse transformation then yields
\begin{align} \label{eq:phisolpos}
    \Phi_>^{(1)}(\omega;\mu) &= e^{V+2\gamma_E a} \int_{0}^\infty \frac{d\omega'}{\omega'} \left(\frac{\mu_0}{\omega'} \right)^{\!a} \Bigg[ G_a\!\left(\frac{\omega}{\omega'}\right) \Phi_>^{(1)}(\omega';\mu_0) \nn \\ &+ \int_0^1 \frac{dx}{1-x} \bigg\{ (x^a -1) G_{a}^{1,0}\!\left(\frac{x\omega}{\omega'} \right) - (x^{-1-a}-x^{-1}) G_{a}^{1,0}\!\left(\frac{\omega}{x\omega'} \right) \bigg\} \Phi_<^{(1)}(-\omega';\mu_0) \nn \\ & - \int_{\mu_0}^\mu \frac{d\mu'}{\mu'} \frac{\alpha_s(\mu') C_F}{\pi} Q_{\rm sp} Q_{M_2}G_{a(\mu,\mu')}^{3,3}\!\left(\frac{\omega}{\omega'}\right) \ln \frac{\mu'}{\mu_0}  \phi_B^+(\omega';\mu_0) \nn \\ &+ \int_0^1 dx \left[ \frac{1}{1-x} \right]_+ \bigg\{ (Q_{\rm sp}^2 + 2 Q_{\rm sp}Q_{M_1} )\bigg(x^a G_a\!\left( \frac{x\omega}{\omega'} \right)+ x^{-a} G_a\!\left( \frac{\omega}{x\omega'} \right)  \bigg) \ln \frac{\mu}{\mu_0} \nn \\ &
    - Q_{\rm sp} Q_{M_1} \bigg( h(x)G_a\left( \frac{x\omega}{\omega'} \right) + h(x^{-1})G_a\!\left( \frac{\omega}{x\omega'} \right) \bigg) \nn \\ 
    &-Q_{\rm sp} Q_{M_2}\bigg( - h(x)G_a\!\left( \frac{x\omega}{\omega'} \right) + x^{-1} h(x^{-1})G_a\!\left( \frac{\omega}{x\omega'} \right) \bigg) \bigg\} \,\phi_B^+(\omega';\mu_0) \nn \\ 
    &+ \bigg\{ -(Q_{\rm sp}^2 + 2 Q_{\rm sp}Q_{M_1} )\left(\frac12 \ln^2 \frac{\mu}{\mu_0} + \ln \frac{\mu_0}{\omega'} \ln \frac{\mu}{\mu_0} \right)  \nn \\ &+ \left( \frac34 Q_{\rm sp}^2 + \frac12 Q_d^2 -i\pi Q_{M_1} Q_{M_2} \right) \ln \frac{\mu}{\mu_0}  \bigg\} G_a\!\left( \frac{\omega}{\omega'} \right) \phi_B^+(\omega';\mu_0) \Bigg] \; , \\ 
    \Phi_<^{(1)}(\omega;\mu) &= e^{V+2\gamma_E a} \int_{0}^\infty \frac{d\omega'}{\omega'} \left(\frac{\mu_0 }{\omega'} \right)^{\!a} \bigg[ G^{2,0}_a\!\left(\frac{-\omega}{\omega'}\right) \Phi_<^{(1)}(-\omega';\mu_0) \nn \\ &- Q_{\rm sp} Q_{M_2} G^{2,1}_a\!\left(\frac{-\omega}{\omega'}\right) \ln \frac{\mu}{\mu_0} \phi_B^+(\omega';\mu_0) \bigg] \; . \label{eq:phisolneg}
\end{align}
In the above, we defined
\begin{align}
    G^{m,n}_a\! \left( \frac{\omega}{\omega'}\right) &\equiv G^{m,n}_{2,2}\! \left( { -a,\ 1-a \atop 1,\ 0  } \bigg|\frac{\omega}{\omega'}\right) \;, \nn \\
    G^{m,n}_{a(\mu,\mu')}\! \left( \frac{\omega}{\omega'}\right) &\equiv G^{m,n}_{4,4}\!\left( { -a(\mu,\mu') ,\ -a(\mu,\mu') ,\ -a,\ 1-a \atop  -a(\mu,\mu') ,\ -a(\mu,\mu') ,\ 1, \ 0  } \bigg|\frac{\omega}{\omega'}\right) \; ,
\end{align}
where $G^{m,n}_{p,q}$ is the general Meijer-G function given in \eqref{eq:genmeijerG}. 
Although to obtain the complete result we have to add the QCD solution in \eqref{eq:alemexpansion}, we will refer to \eqref{eq:phisolpos} and \eqref{eq:phisolneg} as the first-order solution of $\Phi_{B,\qboth}$. For practical applications, this first-order solution will be sufficient to quantify the soft QED effects in the two-body decays $\bar{B} \to M_1 M_2$.


\section{Numerical results} 
\label{sec:numerics}

In order to display the effect of QED evolution due to the 
one-loop anomalous dimension, which corresponds to summing 
QED effects in the 
leading-logarithmic (LL) approximation,\footnote{This terminology 
refers to the one used for the resummation of series 
with one logarithm per loop. The anomalous dimension of the soft 
function as well as of the $B$-meson LCDA in QCD contains 
a cusp term, which produces double logarithms.} we expand the 
initial condition for the soft function at $\mu=\mu_0$ according 
to \eqref{eq:alemexpansion}
\begin{align}\label{eq:inexp}
    \Phi_{B,\qboth} (\omega; \mu_0) = \phi_B^+(\omega;\mu_0) + \frac{\alem}{\pi} \Phi_{B,\qboth}^{(1)}(\omega;\mu_0) + \mathcal{O}(\alem^2) \; .
\end{align}
Within the LL approximation, it is consistent (but not 
necessary) to neglect the QED effect at the initial scale, 
which is of higher order. We therefore set 
$\Phi^{(1)}(\omega;\mu_0)=0$, and adopt the 
exponential model 
\begin{equation} \label{eq:expmodel}
    \phi_B^+ (\omega; \mu_0) = \theta(\omega) \frac{\omega}{\omega_0^2} e^{-\frac{\omega}{\omega_0}}
\end{equation}
for the QCD initial condition at the scale $\mu_0=1$~GeV. 
We set $\omega_0 = 0.3$ GeV, and use $\alpha_s(\mu_0)=0.48$ and fixed $\alem(\mu)=1/134$. We evolve $\alpha_s$ in the $n_f=4$ scheme in the one-loop 
approximation, but neglect the small 
effect from flavour thresholds on the QCD and QED evolution. 

\begin{figure}[p]
    \centering
    \includegraphics[width=0.8\textwidth]{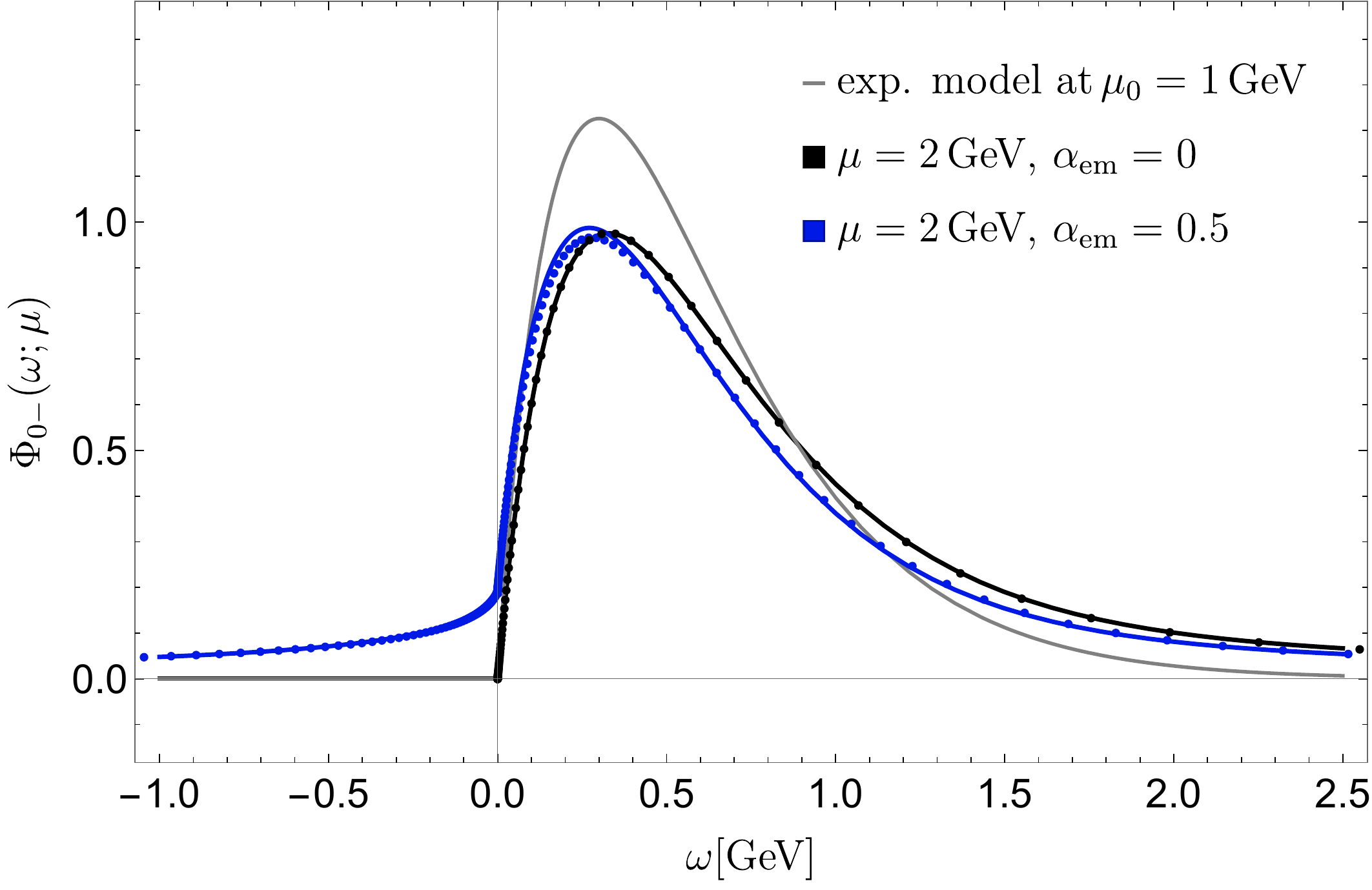} \\[0.7cm] 
\includegraphics[width=0.8\textwidth]{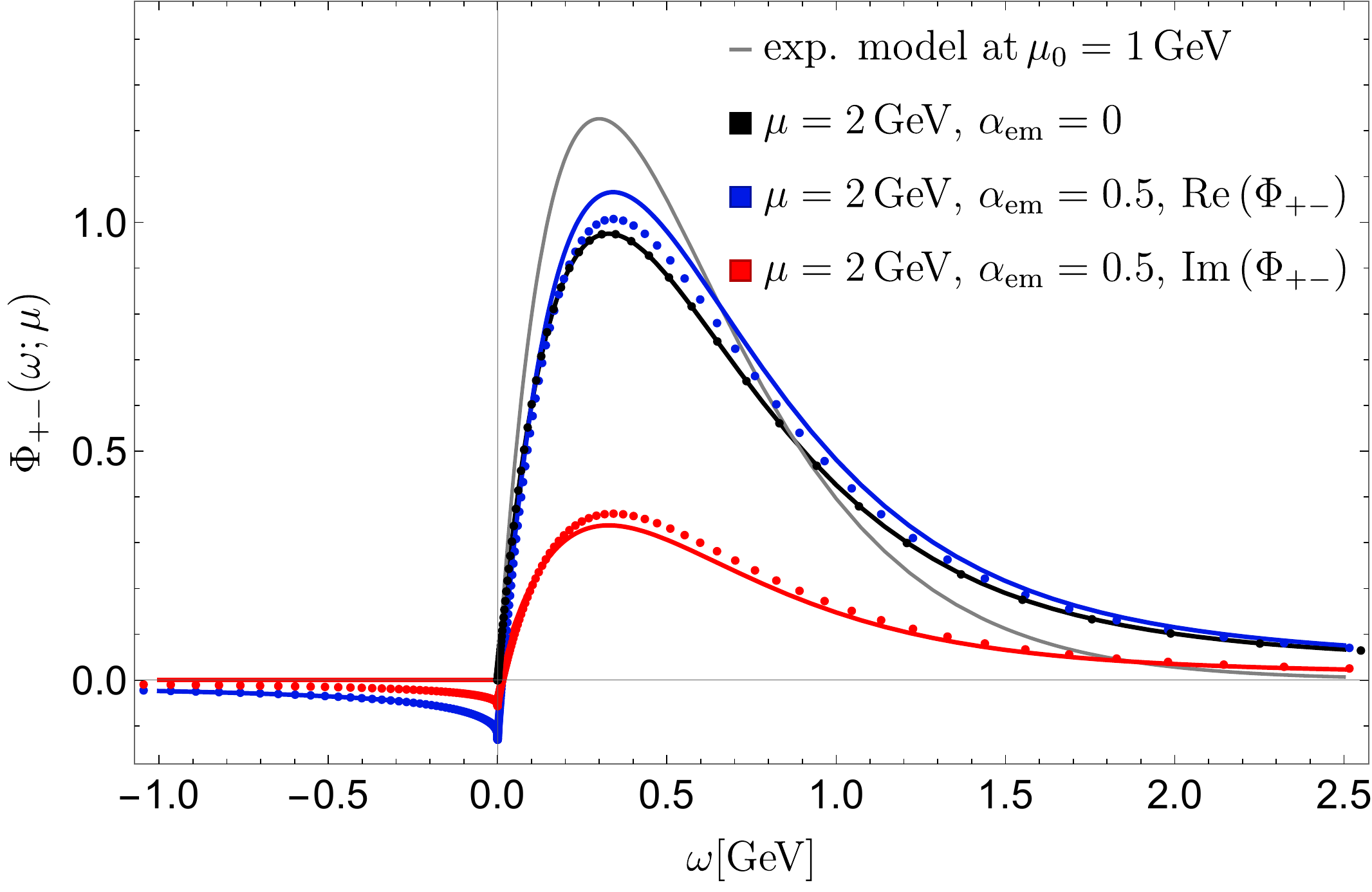}\\[0.2cm]
    \caption{Scale evolution of the soft functions $\Phi_{0-}(\omega;\mu)$ (upper panel) and  $\Phi_{+-}(\omega;\mu)$ (lower panel) in QCD versus QCD$\times$QED. The exponential model~\eqref{eq:expmodel} (gray curve) with $\omega_0 = 0.3$ GeV is evolved from the hadronic scale $\mu_0 = 1$ GeV to a typical hard-collinear scale \mbox{$\mu = 2$~GeV}. The black line shows the analytic result~\eqref{eq:QCDanalyticsol} for the evolved function in the absence of QED effects. The blue (red) curve displays the first-order $\mathcal{O}(\alem)$ solution~\eqref{eq:firstorderexp}, \eqref{eq:firstorderexpIm} for the real (imaginary) part of $\Phi_{+-}(\omega;\mu)$ for $\alem=0.5$. The dots represent the results from the numerical solution obtained by discretization of the unapproximated evolution equation. We recall that 
$\Phi_{0-}(\omega;\mu)$ has zero imaginary part and do not display the corresponding red curve.} 
    \label{fig:0minus}
    \label{fig:plusminus}
\end{figure}

For comparison with our analytic approximations, we solve the 
integro-differential equation  \eqref{eq:B-RGE} numerically. 
The discretization of the RGE  for $\omega \in [-\infty,\infty]$ is obtained with a similar method as in \cite{Beneke:2021pkl}. We introduce an upper cutoff on $\omega$ of $\Omega = 10^4$ GeV and, since the distributions are divergent for $\omega \to 0$, a lower cutoff $\eps = 10^{-7}$. We then discretize the two $\omega$ intervals $[-\eps, - \Omega]$  and $[\eps, \Omega]$ into $N+1$ points each, chosen to accumulate logarithmically towards the boundary values $|\omega_1|=\eps$. The $\omega'$-integral in the differential equation \eqref{eq:B-RGE} can then  be expressed in terms of a Riemann sum. To reduce the integration error the trapezoidal rule is employed, for which the error is roughly proportional to $(\Delta \omega)^3$, where $\Delta \omega$ is the distance between two points. The RGE then turns into a system of $2(N+1)$ coupled first-order differential equations, which can be solved numerically. For $Q_{M_2}=0$ the soft functions are restricted to $\omega>0$ and we only consider the $N+1$ coupled equations obtained by the discretization of the interval $[\epsilon,\Omega]$. For $N=1001$ points, we reproduce the known analytic LL QCD result with an accuracy of $0.02\%$.  The numerical result for the soft function evolved to the scale $\mu=2$~GeV for the cases $\qboth=(0,-)$ and $\qboth=(+,-)$ in QCD$\times$QED is depicted in Fig.~\ref{fig:0minus}. To illustrate the qualitative features of the solution, that is the generation of support at $\omega<0$ and an imaginary part for the case $(+,-)$ of the charged final-state particle, we inflated the electromagnetic coupling to $\alem=0.5$. The real and imaginary part obtained from this numerical solution are shown as blue and red dots, respectively. For comparison, we also show the initial condition (grey) and the evolved result 
in QCD (black), that is for $\alem=0$. The latter follows from \eqref{eq:expanalytic}, and is given by 
\begin{align} \label{eq:QCDanalyticsol}
    \phi_B^+(\omega;\mu) = \frac{1}{\omega_0} e^{V+2\gamma_E a}  \left(\frac{\mu_0}{\omega_0}  \right)^{\!a} F_a\!\left(\frac{\omega}{\omega_0} \right) \; ,
\end{align}
where $ F_a(z) \equiv z\Gamma(2+a) \; _1 F_1\left(2+a, 2 , -z \right)$. (Here and until~\eqref{eq:G3334} the evolution variables $V$ and $a$ refer to the QCD-only definition \eqref{eq:VaQCD}.) 


The numerical solution by discretization can be compared to the first order solution \eqref{eq:phisolpos}, \eqref{eq:phisolneg}. Plugging in the exponential model in \eqref{eq:expmodel} and recalling that $\Phi^{(1)}(\omega;\mu_0)=0$, the first-order solution is obtained by performing explicitly the $\omega'$-integrals \eqref{eq:phisolpos}, \eqref{eq:phisolneg}, using the properties of the Meijer-G functions listed in Appendix~\ref{app:MeijerG}. This can be done either directly or by going back to the Laplace space expressions \eqref{eq:posMellin} and \eqref{eq:QM2MellinsolNeg} with the Laplace transformed initial condition $\tilde{\phi}_B^+(\eta;\mu_0)$. The result is 
\begin{align} 
\label{eq:firstorderexp}
    \Phi_{B,\qboth}^{>(1)}(\omega, \mu) &=  \frac{1}{\omega_0} e^{V+2\gamma_E a} \left( \frac{\mu_0}{\omega_0} \right)^{\!a} \Bigg[-Q_{\rm sp} Q_{M_2} \int_{\mu_0}^{\mu} \frac{d\mu'}{\mu'} \frac{\alpha_s(\mu') C_F}{\pi} G^{33}_{34}\! \left( \frac{\omega}{\omega_0} \right) \ln \frac{\mu'}{\mu_0} \nonumber \\
    &+\int_0^1 dx \left[ \frac{1}{1-x}\right]_+ \bigg(- Q_{\rm sp} Q_{M_1} \bigg\{ h(x) F_a\! \left(\frac{x\omega}{\omega_0}\right) +  h(x^{-1}) F_a\! \left(\frac{\omega}{x\omega_0}\right)\bigg\} \nn \\ 
    &+ Q_{\rm sp} Q_{M_2} \bigg\{ h(x) F_a\!\left( \frac{x\omega}{\omega_0} \right) -x^{-1} h(x^{-1}) F_a\! \left(\frac{\omega}{x\omega_0}\right) \bigg\} \nn \\
    &+ (Q_{\rm sp}^2+2 Q_{\rm sp} Q_{M_1}) x^a F_a\! \left(\frac{x\omega}{\omega_0}\right)\ln \frac{\mu}{\mu_0} \bigg) \nn\\ 
    &+\bigg\{ -(Q_{\rm sp}^2+2 Q_{\rm sp} Q_{M_1} ) \left(\frac{1}{2} \ln^2 \frac{\mu}{\mu_0} + \ln \frac{\mu_0 e^{\gamma_E}}{\omega_0} \ln \frac{\mu}{\mu_0}  \right) \nn \\
    &+ \left( \frac34 Q_{\rm sp}^2 + \frac12 Q_d^2 - i\pi Q_{M_1}Q_{M_2} \right) \ln \frac{\mu}{\mu_0} \bigg\}  F_a\!\left( \frac{\omega}{\omega_0} \right)   \Bigg] \; ,  \\
    \Phi_{B,\qboth}^{<(1)}(\omega,\mu) &= -\frac{Q_{\rm sp} Q_{M_2}}{\omega_0}  e^{V+2\gamma_E a} \left( \frac{\mu_0}{\omega_0} \right)^{\!a} \Gamma(1+a) \Gamma(2+a) \;U\!\left(1+a,0,-\frac{\omega}{\omega_0}\right) \ln \frac{\mu}{\mu_0} \; , 
\label{eq:firstorderexpIm}
\end{align}
where $U(a,b,z)$ is the Tricomi confluent hypergeometric function 
\begin{equation}
    U(a,b,z) = \frac{\Gamma(b-1)}{\Gamma(a)} z^{1-b} \ _1 F_1(a-b+1,2-b,z) + \frac{\Gamma(1-b)}{\Gamma(a-b+1)} \ _1 F_1 (a,b,z) \; ,
\end{equation}
and 
\begin{align}\label{eq:G3334}
    G^{33}_{34} (z) \equiv G^{33}_{34} \bigg( {-a(\mu,\mu'),\ -a(\mu,\mu'), \ -a \atop -a(\mu,\mu'),\ -a(\mu,\mu'), \ 1,\  0} \bigg|z \bigg) \; .
\end{align}
The first-order (in QED, but all-order in QCD) solutions 
\eqref{eq:firstorderexp}, \eqref{eq:firstorderexpIm} are 
shown in solid blue (real part) and red (imaginary part) in 
Fig.~\ref{fig:0minus} and show very good agreement with 
the full numerical result, even for the illustrative, 
large value of $\alem=0.5$.

Returning to the real world with $\alem=1/134$, we focus on 
the inverse-logarithmic moments \eqref{eq:momentdef}, which are relevant, for example, for the non-leptonic decay amplitudes. Particularly important are the moments $\lambda_B$ and $\sigma_{1}$. For the exponential model, their initial values are given by 
\begin{align}
    \lambda_B^{-1} (\mu_0) = \frac{1}{\omega_0}  \; , \qquad \sigma_n(\mu_0) = \sum_{k=0}^n \left( n \atop k \right) (-1)^k \,\Gamma^{(k)}(1) \ln^{n-k}\frac{\mut}{\omega_0} \; ,
\end{align}
where $\Gamma^{(k)}(1) = \partial_z^k \Gamma(z) \big|_{z=1}$. Following \cite{Beneke:2018wjp}, we take $\tilde\mu = \omega_0 e^{-\gamma_E}$ GeV, such that $\sigma_1(\mu_0)=0$. The all-order solutions  \eqref{eq:lambdasol}, \eqref{eq:sigma1sol} for $\lambda_B$ and $\sigma_1$ with  leading-logarithmic QED effects from evolution for the exponential model read
\begin{align}
    \lambda_B^{-1}(\mu) =& e^{V+2\gamma_E a} \frac{\Gamma(1+a)}{\omega_0} \bigg(\frac{\mu_0}{\omega_0} \bigg)^{\!a} \mathcal{F}(0;\mu,\mu_0) \;, \label{eq:momentsexp}\\ 
     \sigma_1(\mu) =& H_\aqed + \gamma_E + \ln \frac{\mut}{\omega_0}  +\int_{\mu_0}^{\mu} \frac{d\mu'}{\mu'}  \frac{\alem(\mu') Q_{\rm sp} Q_{M_1}}{\pi}  \big\{H'_{\aqed(\mu,\mu')} -H'_{-\aqed(\mu,\mu')}  \big\} \; , \label{eq:momentsexp2}
\end{align}
where $a$, $V$ and $\mathcal{F}$ are defined in~\eqref{eq:VQED} and \eqref{eq:calF}, respectively. The numerical values for the various charge combinations at $\mu=2$ GeV are given in Table~\ref{tab:momentresults},\footnote{We cross-checked them with the corresponding values obtained by the discretization method. The numerical solution reproduces the analytical results with permille accuracy, the magnitude of the estimated numerical error.} which shows that the QED correction to the renormalization of the two-body $B \to M_1 M_2$ soft functions is $1.2\%$ at most. This is smaller than the two-loop QCD evolution effect\cite{Braun:2019wyx}. 

\begin{table}[t]\centering 
\begin{tabular}{c|c|c|c|c|c|c}
& $\mu_0=1$ GeV  & \multicolumn{5}{c}{$\mu=2$ GeV}\\ \hline
& initial & QCD & $(0,0)$ & $(-,0)$ & $(0,-)$ & $(+,-)$ \\ \hline \hline
  $\lambda_B^{-1}$ &$3.3333$ &$\phantom{-}2.7922$ &$\phantom{-}2.7918$ & $\phantom{-}2.8018$& $2.7900 +0.0096i$& $2.7977 + 0.0096i$ \\ \hline
  $\sigma_1$ & $0$  &$-0.2125$ & $-0.2129$ &$-0.2099$ & $-0.2139$& $-0.2108$
\end{tabular}
\caption{Numerical values of $\lambda_B^{-1}$ (in units of GeV$^{-1}$) and $\sigma_1$ with $\mut=\omega_0 e^{-\gamma_E}$ and $\omega_0=0.3$ GeV for the exponential model. }
\label{tab:momentresults}
\end{table}

\section{Conclusion}
\label{sec:conclusion}

The present paper continued the investigation of the leading QED effects on exclusive two-body $B$-meson decays through the study of the relevant soft functions  $\Phi_{B,\otimes}(\omega)$.  These were first introduced in \cite{Beneke:2019slt} for the study of $B_s\to \mu^+\mu^-$ and can be viewed as the QED generalization of the leading-twist $B$-meson LCDA $\phi_B^+(\omega)$ for the case of two energetic charged back-to-back final-state particles. However, due to their dependence on the kinematics of two-body decay and the electric charges of the final-state particles, and due to the appearance of rescattering phases, one should rather think of these objects as soft functions relevant in a given process, instead of a universal LCDA.

We computed the $\mathcal{O}(\alem)$ anomalous dimensions for all four possible charge combinations $\otimes = \{(0,0),(-,0),(0,-),(+,-)\}$ of the two-particle final state, and discussed the various cases in detail. While the soft functions for electrically neutral $M_2$ 
are fairly standard generalization of the QCD anomalous dimensions, 
the case of  $Q_{M_2}\not= 0$, i.e.~once soft photons couple to the charged anti-collinear $M_2$ meson, exhibits interesting new features.\footnote{We recall that $M_2$ is the meson that does {\em not} pick up 
the spectator quark from the $B$-meson.}
Among them is the at first sight surprising fact that the support region of $\Phi_{B,\otimes}(\omega)$ must be extended to the entire real axis, $\omega \in (-\infty, \infty)$. The evolution kernels can no longer be expressed in terms of standard plus-distributions, and instead contain the more complicated modified distributions $\oplus$ and $\ominus$. An important consequence is that the first inverse moment $\lambda_B^{-1}$ (but not its logarithmic modifications $\sigma_n$) acquires an imaginary part if $M_2$ is charged, and hence complex phases are generated already from tree-level spectator scattering.

We solved the one-loop evolution equation in Laplace space exactly, i.e.~to all orders in perturbation theory. Furthermore, we solved the RGE in momentum space numerically and derived an analytic solution for the soft functions at first order $\mathcal{O}(\alem)$, that resums large QCD logarithms on top of a fixed-order expansion in $\alem$, which is sufficient for practical purposes. Restricting to the test-function space of inverse-logarithmic moments drastically simplifies the structure of the RGEs and allows us to solve the evolution equations for the coupled system $(\lambda_B^{-1}(\mu),\sigma_n(\mu))$ in QCD$\times$QED analytically to all orders in both coupling constants. Based on these results we obtained numerical estimates of QED effects for the first inverse moment $\lambda_B^{-1}(\mu)$ and the first inverse-logarithmic moment $\sigma_1(\mu)$. When evolved from the hadronic scale $\mu_0=1$ GeV to a typical hard-collinear scale $\mu=2$ GeV, we find QED effects to be at the percent level for $\sigma_1$ and at the permille level for $\lambda_B^{-1}$. 

Despite being numerically small, the findings of this paper are of conceptual importance, and show the consistency of the treatment of QED effects. Together with~\cite{Beneke:2020vnb,Beneke:2021pkl,Beneke:2021jhp}, this work completes the analysis of QED factorization in non-leptonic $B$ decays at scales greater than a few times $\Lambda_{\rm QCD}$ at the one-loop level. Below these scales, the soft functions discussed in this article together with the light-meson LCDAs studied in~\cite{Beneke:2021pkl} need to be matched non-perturbatively onto an effective theory of ultrasoft photons coupling to infinitely heavy, but boosted ultra-relativistic point-like mesons. (In the case of a final-state lepton instead of meson, the matching is of course perturbative.) A better understanding of this matching would complete the theoretical treatment of QED effects in exclusive $B$-meson decays and is hence a desirable goal for future work.

\subsubsection*{Acknowledgements}
We would like to thank Yao Ji for discussions. This research was supported in part by the Deutsche
Forschungsgemeinschaft (DFG, German Research Foundation) through
the Sino-German Collaborative Research Center TRR110 ``Symmetries
and the Emergence of Structure in QCD''
(DFG Project-ID 196253076, NSFC Grant No. 12070131001, -- TRR 110), 
and the Cluster of Excellence PRISMA$^+$ funded by the German Research Foundation (DFG) within
the German Excellence Strategy (Project ID 39083149). J.-N. T. would like to thank
the Studienstiftung des deutschen Volkes for a scholarship.

\appendix

\section{Soft Wilson lines and soft rearrangement}
\label{app:soft}

In this appendix we collect some expressions from \cite{Beneke:2020vnb}, which are required for the calculation of the partonic matrix elements and the anomalous dimensions in Sec.~\ref{sec:detcal}. The soft Wilson lines for outgoing anti-quarks with electric charge $Q_q$ are defined in QCD$\times$QED as 
\begin{equation}
\label{eq:softwilsonlines}
 S^{(q)}_{n_\pm}(x) = \exp \left\{ -i Q_q e \int_0^{\infty} d s \, n_\pm A_{s}(x + s n_\pm) \right\}  \, 
 {\mathbf P} \exp \left\{ -i g_s \int_0^{\infty} d s' \, n_\pm G_s(x + s' n_\pm) \right\} \, .
\end{equation}
For outgoing quarks, $S^{\dagger(q)}_{n_\pm}(x)$ applies instead. The finite-distance Wilson line in \eqref{eq:BLCDAdef} can be written in terms of
\begin{equation}\label{eq:finiteWL}
[tn_-,0]^{(q)} = S_{n_-}^{(Q_q)}(tn_-) 
S_{n_-}^{\dagger (Q_q)}(0)  \; .
\end{equation}
The QCD expression $[tn_-,0]$ is obtained by setting $Q_q=0$ in \eqref{eq:softwilsonlines}. The existence of electrically charged particles in the final state leads to the appearance of outgoing Wilson lines $S_{n_-}^{\dagger (Q_{M_1})}(0) S_{n_+}^{\dagger (Q_{M_2})}(0) $ in the soft operator, which do not decouple and contain the physics of soft rescattering phases. We recall from the main text that this makes the soft function of two-body $B$-decays fundamentally different from the QCD $B$-LCDA. The soft Wilson line associated with a charged meson is defined by
\begin{equation}
    S^{(Q_M)}_{n_\pm}(x) \equiv \left(S_{n_\pm}^{(d)} S_{n_\pm}^{\dagger(u)}\right)(x) = \exp \left\{ -i Q_M e \int_0^{\infty} d s \, n_\pm A_{s}(x + s n_\pm) \right\} \; .
\end{equation}
The QCD part of this Wilson line acts trivially on the colour-neutral states and is therefore equal to unity. Finally, as discussed in detail in \cite{Beneke:2019slt,Beneke:2020vnb} (see also \cite{Beneke:2021pkl}), we introduced the soft rearrangement factors  $R_c$, $R_{\bar{c}}$ in order to consistently define a RGE for the QED-generalization of the light-meson and $B$-meson LCDA. They are defined through the vacuum matrix element
\begin{equation}
\left| \langle 0 |\big(S_{n_-}^{\dagger (Q_{M})} S_{n_+}^{(Q_{M})}\big)(0)\,|0\rangle\right| \equiv R_{c}^{(Q_{M})}  R_{\bar{c}}^{(Q_{M})} \;,
\label{eq:softsubtraction}
\end{equation}
where the absolute value ensures that all imaginary terms due to soft rescattering phases remain in the soft operator itself. The matrix element in \eqref{eq:softsubtraction} is then split into $R_{c (\bar{c})}$ such that the UV divergences only depend on their respective IR regulator $\delta_{c(\bar{c})}$, see Appendix A of \cite{Beneke:2021pkl} for more details. For completeness, we provide their one-loop expressions 
\begin{align}
R_{{c}}^{(Q_{M})} & =  1 -  \frac{\alem}{4\pi} Q_{M}^{\,2} \left[\frac{1}{\epsilon^2} + \frac{2}{\epsilon}\ln{\frac{\mu}{-\delta_{ c}}}  + \ldots \right] , \\
R_{\bar{c}}^{(Q_{M})} & = 1 -  \frac{\alem}{4\pi} Q_{M}^{\,2} \left[\frac{1}{\epsilon^2} + \frac{2}{\epsilon}\ln{\frac{\mu}{-\delta_{\bar c}}} + \ldots \right]  ,
\end{align}
where the dots indicate finite terms and we assume $\delta_{c(\bar{c})} <0$. 


\section{\boldmath Distributions in $\omega$} 
\label{app:wdist}

In the main text, we defined the distributions \eqref{eq:PDwprime} and \eqref{eq:PDwprimeoplusminus} in the variable $\omega'$. Here, we give the anomalous dimensions using plus-distributions in the variable $\omega$, which are relevant for the calculation of the first inverse and logarithmic moments of the soft functions in Sec.~\ref{sec:momentRGEs}.

The plus-distributions in $\omega'$ appearing in \eqref{eq:FGfun} can be directly rewritten in terms of distributions in $\omega$. To this end, we define the (generalized) plus-distribution in $\omega$ analogously to \eqref{eq:PDwprime} and \eqref{eq:PDwprimeoplusminus} by
\begin{align} 
 \int_{-\infty}^\infty d\omega \big[ \dots \big]^{(\omega)}_+ f(\omega) &= \int_{-\infty}^\infty d\omega \big[ \dots \big] (f(\omega)-f(\omega'))\; , \nonumber \\
 \int_{-\infty}^\infty d\omega \big[ \dots \big]^{(\omega)}_{\oplus / \ominus} f(\omega) &=  \int_{-\infty}^\infty d\omega \big[ \dots \big] (f(\omega)-\theta(\pm \omega) f(\omega')) \; . 
\label{eq:PDw}
\end{align}
The relations between the distributions in the different variables are 
\begin{align} \label{eq:wwprimediff}
    \begin{aligned}
     \theta(\omega) \omega' \left[ \frac{\theta(\omega'-\omega)}{\omega'(\omega'-\omega)} \right]_+  &= \theta(\omega')\left[ \frac{\theta(\omega'-\omega)}{\omega'-\omega} \right]^{(\omega)}_{\oplus} - \frac{\theta(-\omega)\theta(\omega')}{\omega'-\omega} \; , \\
     \theta(-\omega) \left[\frac{\theta(\omega'-\omega)}{\omega'-\omega} \right]_\ominus &= \theta(-\omega') \omega \left[ \frac{\theta(\omega'-\omega)}{\omega(\omega'-\omega)} \right]^{(\omega)}_+ + \frac{\theta(-\omega)\theta(\omega')}{\omega'-\omega} \; , \\
      \theta(\omega) \omega \left[ \frac{\theta(\omega'-\omega)}{\omega'(\omega'-\omega)} \right]_+ &=\theta(\omega') \omega \left[\frac{\theta(\omega'-\omega)\theta(\omega)}{\omega'(\omega'-\omega)} \right]^{(\omega)}_+  \; , \\
     \theta(-\omega) \omega \left[ \frac{\theta(\omega-\omega')}{\omega'(\omega-\omega')} \right]_+ &= \theta(-\omega') \omega \left[ \frac{\theta(\omega-\omega')\theta(-\omega)}{\omega'(\omega-\omega')} \right]^{(\omega)}_+  \; , \\
    \theta(\omega) \left[ \frac{\theta(\omega-\omega')}{\omega-\omega'}\right]_\oplus &= \theta(\omega') \omega \left[ \frac{\theta(\omega-\omega')}{\omega(\omega-\omega')} \right]^{(\omega)}_+ + \frac{\theta(\omega)\theta(-\omega')}{\omega-\omega'} \; , \\
     \theta(-\omega) \omega' \left[ \frac{\theta(\omega-\omega')}{\omega'(\omega-\omega')} \right]_+ &= \theta(-\omega') \left[ \frac{\theta(\omega-\omega')}{\omega-\omega'}\right]^{(\omega)}_{\ominus} - \frac{\theta(\omega) \theta(-\omega')}{\omega-\omega'} \; . \end{aligned}
\end{align}
The equal sign is understood in the sense of distributions, meaning that equality holds after integrating against test functions in both $\omega$ and $\omega'$. We note that the terms proportional to $\theta(-\omega)\theta(\omega')$ or $\theta(\omega)\theta(-\omega')$, cancel in the results for the individual diagrams in Fig.~\ref{fig:BLCDAdia} and in the anomalous dimension. 

\subsection{Renormalization factor and anomalous dimension}
The renormalization factors and the anomalous dimensions given in Sec.~\ref{sec:detcal}, can now easily be rewritten in terms of distributions in $\omega$ with the help of \eqref{eq:wwprimediff}. For the anomalous dimension, we find
\begin{align} 
    \Gamma^{(\omega)}_\qboth(\omega,\omega')  &= \frac{\alpha_s C_F}{\pi} \bigg[ \left(\ln \frac{\mu}{\omega-i0} - \frac12 \right)\delta(\omega-\omega') -H^{(\omega)}_+(\omega,\omega') \bigg]\nn  \\ &+ \frac{\alem}{\pi} \bigg[ \bigg(  (Q_{\rm sp}^2 +2Q_{\rm sp} Q_{M_1}) \ln \frac{\mu}{\omega-i0} - \frac34 Q_{\rm sp}^2 -\frac12 Q_d^2  \\ & + i\pi (Q_{\rm sp} + Q_{M_1}) Q_{M_2} \bigg) \delta(\omega-\omega') - Q_{\rm sp} Q_d H^{(\omega)}_+(\omega,\omega') + Q_{\rm sp} Q_{M_2} H^{(\omega)}_-(\omega,\omega')   \bigg] \nn \; ,
\end{align}
which can be obtained from \eqref{eq:BAD} by replacing $H\to H^{(\omega)}$. The latter is defined analogously to \eqref{eq:Hdef}:
\begin{align}\label{eq:Hdefomega}
    H^{(\omega)}_\pm(\omega, \omega') \equiv \theta(\pm\omega') F_\omega^{> (<)} (\omega,\omega')+ \theta(\mp\omega') G_\omega^{< (>)}(\omega,\omega')  \; .
\end{align}
The appearing linear combinations of plus-distributions in $\omega$ are 
\begin{align} \label{eq:FGinomega}
    F_{\omega}^>(\omega,\omega') &= \omega \left[\frac{\theta(\omega'-\omega)\theta(\omega)}{\omega'(\omega'-\omega)} \right]^{(\omega)}_+ + \omega \left[\frac{\theta(\omega-\omega')}{\omega(\omega-\omega')} \right]^{(\omega)}_+ \; , \nonumber  \\
    G_{\omega}^>(\omega,\omega') &=  \omega\left[\frac{\theta(\omega'-\omega)\theta(\omega)}{\omega'(\omega'-\omega)} \right]^{(\omega)}_+ + \left[ \frac{\theta(\omega'-\omega)}{\omega'-\omega} \right]^{(\omega)}_{\oplus} -i\pi \delta(\omega-\omega')\; , \nonumber  \\
    F_{\omega}^<(\omega,\omega') &=  \omega \left[ \frac{\theta(\omega-\omega')\theta(-\omega)}{\omega'(\omega-\omega')} \right]^{(\omega)}_+ + \omega \left[\frac{\theta(\omega'-\omega)}{\omega(\omega'-\omega)} \right]^{(\omega)}_+ \; ,  \nonumber \\
    G_{\omega}^<(\omega,\omega') &= \omega \left[\frac{\theta(\omega-\omega')\theta(-\omega)}{\omega'(\omega-\omega')} \right]^{(\omega)}_+ + \left[\frac{\theta(\omega-\omega')}{\omega-\omega'} \right]^{(\omega)}_{\ominus}  +i\pi \delta(\omega-\omega')\; ,
\end{align}
where the superscript $> (<)$ refers to $\omega'>0 \;(\omega'<0)$. Here, the mixing between positive and negative support is generated through the distributions $G_{\omega}^>$ and $G_{\omega}^<$. Hence, the functions \eqref{eq:Hdefomega} follow from the replacements $\theta(\pm \omega)F^{>(<)}\to \theta(\pm \omega')F^{>(<)}_{\omega}$ and $\theta(\pm \omega)G^{>(<)}\to \theta(\pm \omega')G^{>(<)}_{\omega}$ in \eqref{eq:Hdef}. We conclude that the explicit results for the anomalous dimension of the various charge combinations given in \eqref{eq:G00}--\eqref{eq:G+-} can be obtained simply by $F\to F_\omega$ and $G\to G_\omega$. For $Q_{M_2}\neq 0$, the subscript $>$, $<$ on the anomalous dimension then refers to $\omega'>0$ and $\omega'<0$, respectively.

\subsection{Distributions acting on pure powers} 
\label{app:sec:purepowers}

In order to perform the Laplace transform \eqref{eq:mellintrafo}, there are two options. First, integrating the RGE over $\int\frac{d\omega}{\omega} (\frac{\mu}{\omega})^\eta$ requiring distributions in $\omega$ or, second, using distributions in $\omega'$ acting on $(\frac{\mu}{\omega'})^{-\eta}$, obtained by expressing the functions directly in terms of their Laplace transform or integral representation. Although equivalent, both approaches yield different intermediate results. In the case of the integration for inverse moments, we find for distributions in $\omega$ 
\begin{align}\label{eq:B6}
    \theta(\pm\omega')\int_{-\infty}^\infty \frac{d\omega}{\omega-i0}  \left( \frac{\mu}{\omega-i0} \right)^{\!\eta} F_\omega^{>(<)}(\omega,\omega') &=  \theta(\pm\omega')\, \frac{\left( \frac{\mu}{\omega'-i0} \right)^{\!\eta}}{\omega'-i0} \,(-H_\eta -H_{-\eta})  \; ,  \nn \\ 
    \theta(\pm\omega')\int_{-\infty}^\infty \frac{d\omega}{\omega-i0} \left( \frac{\mu}{\omega-i0} \right)^{\!\eta} G_\omega^{>(<)}(\omega,\omega') &=   \theta(\pm\omega') \,\frac{\left( \frac{\mu}{\omega'-i0} \right)^{\!\eta}}{\omega'-i0}\,(-H_\eta -H_{-\eta}) \ ,  \; 
\end{align}
with $-1<{\rm Re}(\eta)<1$. We used the identity $H_{\eta} - H_{-1-\eta} +\pi \cot (\pi \eta)=0$. This implies
\begin{align}\label{eq:Hexppurepower}
\int_{-\infty}^\infty \frac{d\omega}{\omega-i0} \left(\frac{\mu}{\omega-i0}\right)^{\!\eta} H_\pm^{(\omega)}(\omega,\omega') &=  \frac{ \left(\frac{\mu}{\omega'-i0}\right)^\eta}{\omega'-i0}(-H_\eta-H_{-\eta}) \ ,
\end{align}
which shows that for pure powers the difference between $H^{(\omega)}_+-H^{(\omega)}_-$ in ~\eqref{eq:Hexppurepower} vanishes. Taking derivatives with respect to $\eta$, one rederives the expression \eqref{eq:logF} for the logarithmic moments. Equivalently, the differences $F_\omega^>-G_\omega^>$ and $G_\omega^<-F_\omega^<$ vanish on this function space for $\omega'>0$ and $\omega'<0$, respectively. 

Alternatively, one can consider a transformation with $(\frac{\mu}{-\omega})^\eta$ for $\omega<0$, as done in the main text to solve the RGE to first order in $\alem$. This only changes the expressions in \eqref{eq:B6} for the $G_\omega$ distributions by 
\begin{align}
  \int_{-\infty}^\infty \frac{d\omega}{\omega} \bigg[\theta(\omega)\left(\frac{\mu}{\omega}\right)^{\!\eta} &+\theta(-\omega)\left(\frac{\mu}{-\omega}\right)^{\!\eta}  \bigg] \theta(\pm\omega')G^{>(<)}_\omega(\omega,\omega') \nn \\
  &=   \frac{\theta(\pm\omega')}{\omega'} \left(\frac{\mu}{\pm\omega'}\right)^{\!\eta}(-H_{-\eta}-H_{-1-\eta}-\Gamma(-\eta) \Gamma(1+\eta) \mp i\pi) \; ,
    \end{align}
which holds for $-1< {\rm Re}(\eta) <0$. 
For completeness, we also give the results for the distributions in $\omega'$:
\begin{align}
    \int_{-\infty}^\infty d\omega' \bigg[ \theta(\omega')\left( \frac{\mu}{\omega'} \right)^{\!-\eta}& +\theta(-\omega')\left( \frac{\mu}{-\omega'} \right)^{\!-\eta}\bigg]  \theta(\pm\omega)F^{>(<)}(\omega,\omega') \nn \\ 
    &=  \theta(\pm\omega)\left( \frac{\mu}{\pm\omega} \right)^{\!-\eta} (-H_\eta-H_{-\eta} +\Gamma(-\eta) \Gamma(1+\eta) ) \; , \\
    \int_{-\infty}^\infty d\omega' \bigg[ \theta(\omega')\left( \frac{\mu}{\omega'} \right)^{\!-\eta}& +\theta(-\omega')\left( \frac{\mu}{-\omega'} \right)^{\!-\eta}\bigg]  \theta(\pm\omega)G^{>(<)}(\omega,\omega') \nn \\ 
    &=  \theta(\pm\omega)\left( \frac{\mu}{\pm\omega} \right)^{\!-\eta} (-H_{-\eta}-H_{-1-\eta} \mp i \pi ) \; .
\end{align}

\section{Case of $Q_{M_2}=0$}
\label{sec:asymptoticQM2}
For $Q_{M_2}=0$, the soft function $\Phi_{B,(Q_1,0)}$ can be assumed to only have positive support. The anomalous dimension kernel \eqref{eq:BAD} then reduces to QCD-like expressions. In Laplace space, we obtain
\begin{align}
\label{eq:LaplaceRGEZero}
\left( \frac{d}{d \ln \mu} - \eta \right) \tilde{\Phi}_{B,(Q_1,0)}(\eta;\mu) &= \frac{\alpha_s C_F}{\pi} \bigg[ -H_\eta-H_{-\eta}-\partial_\eta + \frac12 \bigg] \tilde{\Phi}_{B,(Q_1,0)}(\eta;\mu) \nonumber \\ &+ \frac{\alem}{\pi} \bigg[ - (Q_{\rm sp}^2 + 2Q_{\rm sp} Q_{M_1}) \left(  H_\eta + H_{-\eta}+\partial_\eta \right)  \nn \\ & + Q_{\rm sp} Q_{M_1} (H_\eta + H_{-\eta} ) + \frac{3}{4} Q_{\rm sp}^2 + \frac{1}{2} Q_d^2 \bigg]\,\tilde{\Phi}_{B,(Q_1,0)}(\eta;\mu) \; .
\end{align}
Note that this RGE is the same as \eqref{eq:mellinRGE} for $Q_{M_2}=0$. Hence, the solution is
\begin{align} \label{eq:QM2ZeroMellinsol}
\tilde{\Phi}_{B,(Q_1,0)}(\eta;\mu) &= e^{V +2\gamma_E a} \left( \frac{\mu}{\mu_0}\right)^{\!\eta} \frac{\Gamma(1-\eta)}{\Gamma(1+\eta)}\frac{\Gamma(1+\eta+a)}{\Gamma(1-\eta-a)} \mathcal{F}(\eta;\mu,\mu_0 ) \tilde{\Phi}_{B,(Q_1,0)}(\eta+a;\mu_0)  \; ,
\end{align}
which can be also obtained from \eqref{eq:phiallorder}, setting $\tilde{\Phi}_<$ to zero. Expanding this expression to $\mathcal{O}(\alem)$ also reproduces the first-order solution in  \eqref{eq:posMellin} for $Q_{M_2}=0$, provided the initial condition at $\mu=\mu_0$ is expanded according to \eqref{eq:alemexpansion}. 

We study the asymptotic behaviour for the two cases of $i)$ $\omega \to 0$ and $ii)$ $\omega\to \infty$ for \eqref{eq:QM2ZeroMellinsol} in case of the exponential model $\Phi_{B,(Q_1,0)}(\omega;\mu_0) =\omega/\omega_0^2 e^{-\omega/\omega_0}\,\theta(\omega) $. The inverse transformation gives
\begin{align} \label{eq:q1invtrafo}
    \Phi_{B,(Q_1,0)}(\omega;\mu) =& \frac{e^{V+2\gamma_E a}}{\omega_0} \int_{c-i\infty}^{c+i\infty} \frac{d\eta}{2\pi i} \left(\frac{\mu_0}{\omega_0}\right)^{\!a}\left( \frac{\omega}{\omega_0} \right)^{\!\eta} \frac{\Gamma(1-\eta) \Gamma(1+\eta+a)}{\Gamma(1+\eta)} \mathcal{F}(\eta;\mu,\mu_0) \; .
\end{align}
We have $-1<a<0$ and choose $-1-a<c<1+a$. The form of \eqref{eq:q1invtrafo} is equivalent to the soft approximation of the light-meson LCDA near the endpoints \cite{Beneke:2021pkl}. We repeat the analysis in Appendix B.2 of \cite{Beneke:2021pkl} and briefly summarize the results for this particular case.

$i)$ For $\omega \to 0$ we can deform the contour $c\pm i\infty$ to enclose all poles and branch cuts in the right half-plane with respect to ${\rm Re}(\eta)=c$. Therefore, the relevant contribution is obtained from $\Gamma(1-\eta)$ and the leading term in $\omega/\omega_0$ is of the form
\begin{align}
    \Phi_{B,(Q_1,0)}(\omega \to 0;\mu) \sim \frac{\omega}{\omega_0^2} \left(-\ln \frac{\omega}{\omega_0} \right)^{p(\mu)}   \; ,
\label{eq:asymp0}
\end{align}
where $p(\mu)$ is given in \eqref{eq:pdef}. Hence the soft function vanishes linearly for $\omega\to 0$ up to logarithmic corrections, similar to what was observed in~\cite{Beneke:2021pkl}. This result is different from \eqref{eq:phiinf} since the $\eta \to 0$ pole from the all-order solution is absent in this case.

$ii)$ For $\omega \to \infty$, we need to deform the integration contour to enclose the poles and branch cuts on the left half-plane with respect to ${\rm Re}(\eta)=c$. The result is 
\begin{align}
    \Phi_{B,(Q_1,0)}(\omega \to \infty;\mu) \sim \frac{1}{\omega_0} \left( \frac{\omega}{\omega_0} \right)^{\!-1-a} \ln^{p(\mu_0)} \left( \frac{\omega}{\omega_0} \right)  \; ,
\label{eq:winfty}
\end{align}
which agrees to the behaviour of the all-order solution in \eqref{eq:phiinf}.

\section{Properties of Meijer-G functions}
\label{app:MeijerG}
In the main text, we extensively used the representation of the complex line integrals for the inverse Laplace transformation in terms of Meijer-$G$ functions. Here, we briefly recall some of the important properties of these functions from \cite{Luke:1969}. Generally, a Meijer-$G$ function is defined by the complex contour integral 
\begin{align}\label{eq:genmeijerG}
    G^{m,n}_{p,q} \bigg( {\mathbf{a} \atop \mathbf{b} } \bigg| z \bigg) = \int_\mathcal{C} \frac{d\eta}{2\pi i} z^\eta \frac{\prod_{j=1}^{m}\Gamma(b_j-\eta)\prod_{j=1}^{n}\Gamma(1-a_j+\eta)}{\prod_{j=m+1}^{q}\Gamma(1-b_j+\eta) \prod_{j=n+1}^{p}\Gamma(a_j-\eta)} 
\end{align}
for integer $0\leq m \leq q$ and $0\leq n \leq p$, where $\mathbf{a}=(a_1,\dots,a_p)$ and $\mathbf{b}=(b_1,\dots,b_q)$. The contour $\mathcal{C}$ can be chosen in three ways with different convergence properties depending on the analytic structure of the Gamma functions of the integrand. The relevant contour for our purposes is a straight line from $c-i\infty$ to $c+i\infty$ with some real parameter $c$ separating the strings of poles from the Gamma functions in the numerator. In this case, the contour integral \eqref{eq:genmeijerG} converges for $p+q<2(n+m)$ and $|{\rm arg}(z)| < ( n+m - (p+q)/2)\pi$. For $z\to 0$ one may deform the contour $\mathcal{C}$ to a path starting at ${\rm Re}(\eta) = \infty$ that encircles the string of poles from $\Gamma(b_j-\eta)$ in mathematically negative direction. Equivalently, for $z\to \infty$, the contour can be deformed to start at ${\rm Re}(\eta) =-\infty$ and encircle all poles of $\Gamma(1-a_j+\eta)$ in positive direction. This is used in the main text to derive the endpoint behaviour of the soft functions for small and large arguments. For large values of the indices $n,m,p,q$, the Meijer-G functions can be difficult to evaluate numerically. 

To simplify our results, there are two practical relations, namely the closure property with respect to integration over positive arguments and the inversion formula
\begin{align}
    \int_0^\infty dz\ &G^{m,n}_{p,q} \bigg( {\mathbf{a} \atop \mathbf{b} } \bigg| \lambda z \bigg) G^{\mu,\nu}_{\sigma,\tau} \bigg( {\mathbf{c} \atop \mathbf{d}} \bigg| \omega z \bigg) = \frac{1}{\lambda} G^{n+\mu,m+\nu}_{q+\sigma,p+\tau} \bigg( {-b_1,\dots -b_m,\mathbf{c},-b_{m+1} ,\dots,-b_q \atop -a_1,\dots ,-a_n ,\mathbf{d} , -a_{n+1} ,\dots,-a_p} \bigg | \frac{\omega}{\lambda} \bigg)\; , \nn \\
     &G^{m,n}_{p,q} \bigg( {\mathbf{a} \atop \mathbf{b} } \bigg| z \bigg) = G^{n,m}_{q,p} \bigg( {1-\mathbf{a} \atop 1- \mathbf{b} } \bigg| \frac{1}{z} \bigg) \; ,
\end{align}
respectively.  The first equation implies a formula for the Laplace transformation,
\begin{align}
    \int_0^\infty dz~z^{\alpha} e^{-\omega z} G^{m,n}_{p,q} \bigg( {\mathbf{a} \atop \mathbf{b} }\bigg| \lambda z  \bigg) = \omega^{-\alpha-1} G^{m,n+1}_{p+1,q} \bigg( {-\alpha,\ \mathbf{a} \atop \mathbf{b} } \bigg| \frac{\lambda}{\omega} \bigg)  \; .
\end{align}
We used these relations implicitly in Laplace space to simplify the $\omega'$-integrals in the first order solution \eqref{eq:phisolpos} and \eqref{eq:phisolneg} after fixing the initial condition to the exponential model. For certain arguments the Meijer-$G$ functions can become singular at $z=1$. This is also true in our case, specifically for $G_a(z)$ defined in \eqref{eq:QCDMeijerG}. To avoid complications, we need to show that the singularity for $z\to 1$ is integrable as long as the initial condition is a regular function at this point. From the expression of the Meijer-$G$ function \eqref{eq:QCDMeijerG} in terms of the hypergeometric function, we find, following \cite{Liu:2020eqe},
\begin{align}
    \lim_{z\to 1}G_{2,2}^{1,1} \bigg( {{-a,\ 1-a} \atop {1,\ 0}} \bigg| z \bigg) = - \frac{\sin \pi a}{\pi} \frac{\Gamma(1+2a)}{|1-z|^{1+2a}} +\mathcal{O}(1)\; .
\end{align}
In physical applications we have $a<0$, and the singularity is indeed integrable. The limit $z\to1$ can also be analyzed for the other Meijer-$G$ functions defined in the main text. This analysis is more complicated, but in all practically relevant cases the exponent of the singular term is proportional to $|1-z|^{-1-2a}$. Hence, we conclude that all singularities are integrable and $\Phi_{B,\qboth}^{(1)}$ in \eqref{eq:phisolpos},\eqref{eq:phisolneg} is regular. 


\bibliographystyle{JHEP} 
\bibliography{refs.bib}
\end{document}